\documentclass[a4paper,review,11pt,authoryear]{elsarticle}
\usepackage{amsmath,bm,hyperref,url,fullpage,mathtools,booktabs,bbm}
\usepackage{amssymb}
\usepackage{paralist,enumitem,todonotes}
\usepackage{multirow}
\usepackage{threeparttable}
\usepackage{rotating}
\usepackage{float}
\usepackage{graphicx}
\usepackage[ruled,linesnumbered]{algorithm2e}
\usepackage{breakurl}
\usepackage{caption}
\usepackage{subcaption}
\usepackage[a4paper, margin=1in]{geometry}
\usepackage{changes}
\usepackage{enumitem}
\usepackage{algorithm2e}
\usepackage{hyperref}
\hypersetup{
	colorlinks=true,
	linkcolor=blue,
	citecolor=blue,
	urlcolor=blue}
\newtheorem{theorem}{Theorem}  

\newtheorem{lemma}{Lemma}

\newcommand{\bbeta}{\boldsymbol{\beta}}

\begin{document}
\date{}
\begin{frontmatter}
  \title{Debias Online Expectile Regression for High-Dimensional Stream Data and Statistic Inference}
    \author[1]{Wei Cao} 
    \author[1,2]{Shanshan Wang}
    \ead{Corresponding author at (School of Economics and Management, Beihang University, Beijing 100191, China) via sswang@buaa.edu.cn.}
    \address[1]{School of Economics and Management, Beihang
			University, Beijing, China}
	\address[2]{MOE Key Laboratory of Complex System Analysis and Management Decision, Beihang University}
    \begin{abstract}
    Expectile regression is a powerful tool for capturing heterogeneous structures in data and has attracted increasing attention in large-scale data analysis due to its computational efficiency. However, statistical inference for high-dimensional online expectile regression remains underexplored. In this paper, we propose a debiased online expectile regression method for high-dimensional streaming data. The proposed method incorporates Lasso regularization to perform variable selection and alleviate estimation instability caused by high dimensionality. It further adopts an online updating strategy in which recursively updated cumulative statistics are used to approximate the global gradient, and a proximal gradient descent algorithm is employed to update model parameters. To enable valid inference when only the current batch and historical summary statistics are available, we also develop an online precision matrix estimation procedure. We establish the consistency and asymptotic normality of the proposed debiased estimator. Extensive numerical studies and real-data applications show that the proposed method achieves performance comparable to the oracle and offline estimators while maintaining low storage cost.
   
    \end{abstract}
				
		
\end{frontmatter}
	
\section{Introduction}
\label{sec:introduction}

Quantile regression (QR), introduced by \citet{koenker1978regression}, provides a powerful tool for modeling heterogeneous effects across different locations of the conditional distribution. By minimizing an asymmetric absolute loss, QR improves upon mean regression in robustness and distributional characterization. However, the check loss used in QR is non-differentiable at zero and often requires linear programming algorithms for estimation. Expectile regression (ER), developed by \citet{newey1987asymmetric}, offers an alternative distributional regression framework based on an asymmetric squared loss. For $\tau\in(0,1)$, the conditional $\tau$-expectile of $Y$ given $\mathbf X=\boldsymbol{x}$ is defined as
\begin{equation}\label{expectile}
e_\tau(Y\mid \mathbf X=\boldsymbol{x})
=
\mathop{\arg\min}_{\theta}
\mathbb E
\left[
\rho_\tau(Y-\theta)
\mid
\mathbf X=\boldsymbol{x}
\right].
\end{equation}
Compared with QR, ER has a continuously differentiable loss function and can be efficiently solved by iteratively reweighted least squares. It also incorporates both the direction and magnitude of residuals, making it useful for characterizing tail behavior. Theoretical foundations of ER, including its relationship with quantiles, identifiability, consistency, and asymptotic normality, have been established in the literatures \citep{newey1987asymmetric,jones1994expectiles,abdous1995relating}.

Due to these advantages, ER has been widely used in applications involving distributional heterogeneity and tail risk. In finance, expectiles provide a useful basis for risk measurement, including the estimation of Value at Risk and Expected Shortfall \citep{taylor2008estimating,Zigelexpect,bellini2014generalized,bellini2017risk,daouia2018estimation,xu2022prediction}. For real application, ER has also been applied to portfolio allocation and actuarial risk management \citep{garg2026portfolio,cai2019reinsurance}. Recently, ER methods have also been successfully applied to survival analysis, particularly for mean residual life estimation, reflecting the growing interest in expectile-based modeling \citep{seipp_weighted_2021,ciuperca2025right,zhang2024weighted,cao2026expectile,cao2026Neural}.

Above studies focused on low-dimensional settings. However, high-dimensional data are now common in fields such as finance, genomics, and biomedical research, where the number of covariates may be comparable to or even larger than the sample size. In such settings, conventional regression methods often suffer from non-identifiability, multicollinearity, and high computational cost. Regularization methods, such as ridge regression, Lasso, Elastic Net, Adaptive Lasso, and nonconvex penalties, provide standard tools for sparse estimation and variable selection \citep{hoerl1970ridge,Tibshirani1996RegressionSA,zou2005regularization,zou2006adaptive,zhang2010nearly}. 
High-dimensional ER has also received increasing attention. For example, \citet{gu_high-dimensional_2016} studied high-dimensional sparse ER and established theoretical properties under several penalty functions, including Lasso, SCAD, and MCP. Subsequent studies further developed penalized ER methods based on SCAD, Elastic Net, and Adaptive Lasso penalties \citep{zhao2018expectile,liao2019penalized,xu2021elastic,ciuperca2021variable}. Recent studies extend high-dimensional ER to more complex modeling structures. For instance, \citet{pan2023high} incorporated graph structures among predictors into high-dimensional ER to exploit covariate network information, while \citet{soale2023projection} developed expectile regression based on sufficient dimension reduction. Other studies have considered group structures, heavy-tailed errors, heteroscedasticity, and semiparametric modeling \citep{ouhourane2024group,zhao2025semiparametric}. Overall, these studies have substantially advanced high-dimensional ER from sparse variable selection toward more flexible frameworks involving structured modeling, dimension reduction, and complex error distributions.

Despite these developments, statistical inference for high-dimensional ER remains challenging. Penalized estimators are biased due to shrinkage, so statistical inference based on classical low-dimensional theory are no longer directly valid \citep{zhang2008bias}. Debias technique has been developed for high-dimensional linear models \citep[see, e.g.,][]{van2014asymptotically,zhang2014confidence}, and related inference methods have recently been extended to ER \citep{li2024inference,ciuperca2025smoothed}. Nevertheless, these studies mainly consider static datasets where all observations are available simultaneously. In many modern applications, data arrive sequentially in batches. Streaming data are more challenge for high-dimensional inference, where historical raw samples may no longer be accessible. Recent studies have developed online inference methods for high-dimensional generalized linear models and QR \citep{luo2023statistical,han2026adaptive,peng2024two}. However, to the best of our knowledge, high-dimensional online inference for ER has not been systematically investigated.

This paper studies online sparse estimation and statistical inference for high-dimensional streaming ER. We combine Lasso regularization with a renewable estimation framework and develop a \underline{de}biased \underline{c}umulative \underline{o}nline \underline{e}xpectile \underline{r}egression method, for simplicity we term the proposed method as \textbf{DeCOER} in follows. The main contributions are summarized as follows:
\begin{enumerate}[label=(\arabic*)]
    \item We develop a Lasso-regularized online ER estimator for high-dimensional streaming data. By combining proximal gradient descent with recursive cumulative statistics, the proposed method updates the estimator without storing historical raw data and handles the non-differentiability of the $\ell_1$ penalty.

    \item We propose an online debiasing procedure for correcting the shrinkage bias of the Lasso estimator. To estimate the precision matrix in high dimensions, we reformulate nodewise regression using cumulative covariance statistics, making it applicable to streaming data without repeated access to historical samples.

    \item We establish the asymptotic theory for the proposed debiased estimator. Under suitable regularity conditions, we prove its asymptotic normality and construct online confidence intervals and hypothesis tests for high-dimensional streaming ER.
\end{enumerate}

The remainder of this paper is organized as follows. Section~\ref{sec:preliminary} reviews high-dimensional ER in the offline setting. Section~\ref{sec:Method} introduces the proposed debiased cumulative online ER method. Section~\ref{sec:precision} presents the online precision matrix estimation procedure, and Section~\ref{sec:theory} establishes the asymptotic theory and inference procedure. Sections~\ref{sec:simulation} and Section~\ref{sec:empirical} report numerical simulations and empirical studies, respectively. Section~\ref{sec:conclusion} concludes the paper and discusses future research directions.

\section{Preliminary}\label{sec:preliminary}

In this section, we first introduce offline expectile regression for high dimension data. Here, consider independent and identically distributed total sample as $\{\boldsymbol{x}_i,y_i\}_{i=1}^{N}$, where $y_i$ is a scalar variable and $\boldsymbol{x}_i = (x_{i,1}, \ldots, x_{i,p})^\top$ is a $p$-dimensional covariate. And we assume the dimension $p$ is allowed to grow with the sample size or even be much larger than $N$. Assume that data are generated by the following high-dimensional expectile linear model:
\begin{equation*}
    e_{\tau}(y_i|\boldsymbol{x}_i)=\boldsymbol{x}_i^\top\boldsymbol{\beta}_0(\tau),
\end{equation*}
where $\bbeta (\tau)$ is the unknown expectile regression parameter. To simplify notation, we omit $\tau$ when there is no ambiguity. Under high dimensional setting, a prevalent techniques is inducing a regularization term, here we involve the lasso penalty term. Hence the lasso estimator can be derived by: 
\begin{equation}\label{eq:penalty}
\hat{\boldsymbol{\beta}}_\lambda = \mathop{\arg\min}_{\boldsymbol{\beta}} \left( \mathcal{Q}_{N_b}(\boldsymbol{\beta}) + P_\lambda(\boldsymbol{\beta}) \right) = \mathop{\arg\min}_{\boldsymbol{\beta}} \left( \frac{1}{N_b} \sum_{i=1}^{N_b} \rho_\tau(y_i - \boldsymbol{x}_i^\top \boldsymbol{\beta}) + \lambda\|\boldsymbol{\beta}\|_1 \right),
\end{equation}
here $\lambda$ is the regularization  parameter. For simplicity, we reformulate $\mathcal{Q}_{N_b}(\boldsymbol{\beta})$ in Eq.~\eqref{eq:penalty} in matrix form. Due to the ALS of expectile regression we introduce a related weighting matrix  related to $\boldsymbol\beta$, defined as  $\mathbf{W}_{\boldsymbol{\beta}} = \text{diag}(w_{\boldsymbol{\beta},1}, \ldots, w_{\boldsymbol{\beta},N_b})$, where
$w_{\boldsymbol{\beta},i} = \left|\tau - \boldsymbol{I}(y_i < \boldsymbol{x}_i^\top \boldsymbol{\beta})\right|$
is the diagonal element. Consequently, the expectile loss function in ~\eqref{eq:penalty} can be rewritten as follows,
\begin{equation}\label{eq:penaltymatirx}
\mathcal{Q}_{N_b}(\boldsymbol{\beta}) = \frac{1}{2N_b} (\boldsymbol Y - \mathbf{X}\boldsymbol{\beta})^\top \mathbf{W}_{\boldsymbol{\beta}} (\boldsymbol Y - \mathbf{X}\boldsymbol{\beta}),
\end{equation}
here $\boldsymbol Y=(y_1,\ldots,y_{N_b})^\top, \mathbf{X}= (\boldsymbol{x}_1,\ldots,\boldsymbol{x}_{N_b})^\top$. However, it is known that the estimation of $\hat{\boldsymbol{\beta}}_\lambda$ is problematic due to inherent biases inducing of Lasso penalty. 

Under ideal situation, assume the weight matrix $\mathbf{W}_{\boldsymbol{\beta}_0}$ with true parameters $\boldsymbol{\beta}_0$ is known, Eq.\eqref{eq:penaltymatirx} degenerated as optimization problem of Weighted Generalized Least Squares (WGLS). Hence, the debias expectile estimator can be derived under WGLS framework. Specifically, denoted
$\hat{\boldsymbol{\epsilon}} = \boldsymbol Y - \mathbf{X}\hat{\boldsymbol{\beta}}_{\lambda}$ as the estimation of residual term, and replace the unknown weight matrix with $\mathbf{W}_{\hat{\boldsymbol{\beta}}_{\lambda}}$, therefore the debias estimator are: 
\begin{equation}\label{eq:debias}
\hat{\boldsymbol{\beta}}_\text{de}=
\hat{\boldsymbol{\beta}}_{\lambda}
+
\hat{\boldsymbol{\Theta}}
\frac{
\mathbf{X}^\top
\mathbf{W}_{\hat{\boldsymbol{\beta}}_\lambda}
\hat{\boldsymbol{\epsilon}}
}{N_b},
\end{equation}
here,
$\hat{\boldsymbol{\Sigma}}
=
\mathbf X^\top
\mathbf W_{\hat{\boldsymbol{\beta}}_{\lambda}}
\mathbf X/{N_b}$ is the generalized Hessian matrix related to estimation
$\hat{\boldsymbol{\beta}}_\lambda$, $\hat{\boldsymbol{\Theta}}$ is the inverse matrix of Hessian, also termed as precision matrix. Prevalent research, often deduce the precision matrix with nodewise regression.

\section{Debiased Cumulative Online Expectile Regression}\label{sec:Method}

\subsection{Model Framework}\label{sec2-sub1}

When data are generated dynamically in a streaming manner, the full offline dataset
$\{\mathbf{X},\boldsymbol{Y}\}$ is usually not available or cannot be loaded into memory all at once. Therefore, the offline estimation method introduced in Section~\ref{sec:preliminary} cannot be directly applied for sparse parameter estimation. To address this issue, this subsection develops a high-dimensional debiased online estimation model for streaming data.

Specifically, let the $t$-th data batch denoted by
$\mathcal{D}_t=\{\mathbf{X}^{(t)},\boldsymbol{Y}^{(t)}\}$
with sample size $n_t$. The cumulative sample size up to batch $b$ is
$N_b=\sum_{t=1}^b n_t$. In the streaming setting, the full-sample objective function based on the first $b$ batches can be written as
\begin{equation}\label{eq:penalty-stream}
\begin{aligned}
\mathcal Q_{N_b,\lambda}(\boldsymbol{\beta})
={}&
\frac1{2N_b}
\sum_{t=1}^{b-1}
(\boldsymbol Y^{(t)}-\mathbf X^{(t)}\boldsymbol{\beta})^\top
\mathbf W_{\boldsymbol{\beta}}^{(t)}
(\boldsymbol Y^{(t)}-\mathbf X^{(t)}\boldsymbol{\beta})\\
&+
\frac1{2N_b}
(\boldsymbol Y^{(b)}-\mathbf X^{(b)}\boldsymbol{\beta})^\top
\mathbf W_{\boldsymbol{\beta}}^{(b)}
(\boldsymbol Y^{(b)}-\mathbf X^{(b)}\boldsymbol{\beta})
+
\lambda_b\|\boldsymbol{\beta}\|_1 .
\end{aligned}
\end{equation}
Here, $\mathbf{W}^{(t)}_{\boldsymbol{\beta}}$ is the diagonal residual weight matrix for the $t$-th data batch.

Equation~\eqref{eq:penalty-stream} decomposes the full-sample loss into the cumulative contributions of all data batches. If historical raw data were repeatedly accessible, the historical weights
$\mathbf W_{\boldsymbol{\beta}}^{(t)}$ could be updated in real time according to the current parameter $\boldsymbol{\beta}$. However, in streaming data settings, historical batches are typically discarded after being processed. Therefore, in this study we adopt a ``frozen-weight'' approximation for historical batches. Specifically, for $t<b$, we set
$
\mathbf W_{\boldsymbol{\beta}}^{(t)}
\approx
\mathbf W_{\tilde{\boldsymbol{\beta}}_t}^{(t)}
$,
where $\tilde{\boldsymbol{\beta}}_t$ is the online estimator obtained after processing the $t$-th batch. This approximation compresses the historical contribution to the objective function into recursively updated summary statistics, thereby avoiding repeated storage and access to historical raw data. Hence the surrogate empirical loss corresponding to Eq.~\eqref{eq:penalty-stream} is defined as:
\begin{equation}\label{eq:penalty-stream-proxy}
\begin{aligned}
\mathcal Q_{N_b}^{\mathrm{s}}(\boldsymbol{\beta})
={}&
\frac1{2N_b}
\sum_{t=1}^{b-1}
(\boldsymbol Y^{(t)}-\mathbf X^{(t)}\boldsymbol{\beta})^\top
\mathbf W_{\tilde{\boldsymbol{\beta}}_t}^{(t)}
(\boldsymbol Y^{(t)}-\mathbf X^{(t)}\boldsymbol{\beta})\\
&+
\frac1{2N_b}
(\boldsymbol Y^{(b)}-\mathbf X^{(b)}\boldsymbol{\beta})^\top
\mathbf W_{\boldsymbol{\beta}}^{(b)}
(\boldsymbol Y^{(b)}-\mathbf X^{(b)}\boldsymbol{\beta}) .
\end{aligned}
\end{equation}
After expanding the historical part and omitting constants independent of $\boldsymbol{\beta}$, Eq.~\eqref{eq:penalty-stream-proxy} can be equivalently written as
\begin{equation*}\label{eq:DeCOER-proxy-loss}
\mathcal Q_{N_b}^{\mathrm{s}}(\boldsymbol{\beta})
=
\frac{1}{2N_b}
\left\{
\boldsymbol{\beta}^\top\mathbf Z_{N_{b-1}}\boldsymbol{\beta}
-
2\mathbf U_{N_{b-1}}^\top\boldsymbol{\beta}
+
(\boldsymbol Y^{(b)}-\mathbf X^{(b)}\boldsymbol{\beta})^\top
\mathbf W_{\boldsymbol{\beta}}^{(b)}
(\boldsymbol Y^{(b)}-\mathbf X^{(b)}\boldsymbol{\beta})
\right\},
\end{equation*}
where
$\mathbf Z_{N_{b-1}}
=
\sum_{t=1}^{b-1}
\mathbf X^{(t)\top}
\mathbf W_{\tilde{\boldsymbol{\beta}}_t}^{(t)}
\mathbf X^{(t)}$
and
$\mathbf U_{N_{b-1}}
=
\sum_{t=1}^{b-1}
\mathbf X^{(t)\top}
\mathbf W_{\tilde{\boldsymbol{\beta}}_t}^{(t)}
\boldsymbol Y^{(t)}$.
Based on this surrogate loss, the recursive surrogate objective with the Lasso penalty is defined as
\begin{equation*}\label{eq:DeCOER-proxy-objective}
\mathcal Q_{N_b,\lambda}^{\mathrm{s}}(\boldsymbol{\beta})
=
\mathcal Q_{N_b}^{\mathrm{s}}(\boldsymbol{\beta})
+
\lambda_b\|\boldsymbol{\beta}\|_1 .
\end{equation*}

Since the $\ell_1$ penalty is non-differentiable at zero, this study adopts proximal gradient descent (PGD) to solve for $\boldsymbol{\beta}$. Define the gradient vector and Hessian-type matrix of the surrogate loss are:
\begin{equation}\label{eq:DeCOER-proxy-stats}
\begin{aligned}
&\nabla\mathcal Q_{N_b}^{\mathrm{s}}(\boldsymbol{\beta})
=
\frac1{N_b}
\left\{
\mathbf Z_{N_{b-1}}\boldsymbol{\beta}
-
\mathbf U_{N_{b-1}}
+
\mathbf X^{(b)\top}
\mathbf W_{\boldsymbol{\beta}}^{(b)}
(\mathbf X^{(b)}\boldsymbol{\beta}-\boldsymbol Y^{(b)})
\right\},\\
&\nabla^2\mathcal Q_{N_b}^{\mathrm{s}}(\boldsymbol{\beta})
=
\frac1{N_b}
\left\{
\mathbf Z_{N_{b-1}}
+
\mathbf X^{(b)\top}
\mathbf W_{\boldsymbol{\beta}}^{(b)}
\mathbf X^{(b)}
\right\}.
\end{aligned}
\end{equation}

When the $b$-th data batch arrives, PGD iterations are performed within the current batch. Specifically, at the $k$-th update, the current-batch weight matrix is computed based on the previous iterate
$\tilde{\boldsymbol{\beta}}_b^{(k-1)}$ and the cumulative statistics are constructed as:
\begin{equation*}
\begin{aligned}
\mathbf Z_{N_b}^{(k)}
&=
\mathbf Z_{N_{b-1}}
+
\mathbf X^{(b)\top}
\mathbf W_{\tilde{\boldsymbol{\beta}}_b^{(k-1)}}^{(b)}
\mathbf X^{(b)},\\
\mathbf U_{N_b}^{(k)}
&=
\mathbf U_{N_{b-1}}
+
\mathbf X^{(b)\top}
\mathbf W_{\tilde{\boldsymbol{\beta}}_b^{(k-1)}}^{(b)}
\boldsymbol Y^{(b)}.
\end{aligned}
\end{equation*}
Substituting
$\boldsymbol{\beta}=\tilde{\boldsymbol{\beta}}_b^{(k-1)}$
into Eq.~\eqref{eq:DeCOER-proxy-stats}, the approximate gradient used in the $k$-th update is
\begin{equation*}
\boldsymbol g_{N_b}^{(k)}
=
\frac1{N_b}
\left\{
\mathbf Z_{N_b}^{(k)}
\tilde{\boldsymbol{\beta}}_b^{(k-1)}
-
\mathbf U_{N_b}^{(k)}
\right\}.
\end{equation*}
Therefore, the $k$-th proximal gradient update is
\begin{equation*}
\tilde{\boldsymbol{\beta}}_b^{(k)}
=
\operatorname{prox}_{\eta,\lambda_b}
\left[
\tilde{\boldsymbol{\beta}}_b^{(k-1)}
-
\eta\boldsymbol g_{N_b}^{(k)}
\right],
\end{equation*}
where $\eta$ is the step size and
$\operatorname{prox}_{\eta,\lambda_b}$ denotes the soft-thresholding operator, defined as $\operatorname{prox}_{\eta,\lambda}(v)
=
\operatorname{sgn}(v)
\left(|v|-\eta\lambda\right)_+
=
\operatorname{sgn}(v)\max\{|v|-\eta\lambda,0\}$. In practical implementation, the online estimator from the previous batch is used as the initial value for the current batch, $
\tilde{\boldsymbol{\beta}}_b^{(0)}
=
\tilde{\boldsymbol{\beta}}_{b-1}$.
If
$
\|
\tilde{\boldsymbol{\beta}}_b^{(k)}
-
\tilde{\boldsymbol{\beta}}_b^{(k-1)}
\|_{\infty}
<
\varepsilon
$,
the inner iteration for the current batch is regarded as converged; otherwise, the update continues until the maximum number of iterations $K$ is reached.

After convergence within the $b$-th batch, denote the final estimator by
$\tilde{\boldsymbol{\beta}}_b$. The historical cumulative statistics are then updated as
$\mathbf{Z}_{N_b}
=
\mathbf{Z}_{N_{b-1}}
+
\mathbf{X}^{(b)\top}
\mathbf{W}^{(b)}_{\tilde{\boldsymbol{\beta}}_b}
\mathbf{X}^{(b)}$
and
$\mathbf{U}_{N_b}
=
\mathbf{U}_{N_{b-1}}
+
\mathbf{X}^{(b)\top}
\mathbf{W}^{(b)}_{\tilde{\boldsymbol{\beta}}_b}
\mathbf{Y}^{(b)}$.
The updated $\mathbf{Z}_{N_b}$ and $\mathbf{U}_{N_b}$ serve as historical summaries when batch $b+1$ arrives. Algorithm~\ref{alg:local_pgd} summarizes the local PGD iterative update procedure when the $b$-th data batch arrives. Specifically, the algorithm uses the updated estimator from the previous batch,
$\tilde{\boldsymbol{\beta}}_{b-1}$, as the initial value for the current batch. At the $k$-th iteration, the approximate gradient is computed from the historical cumulative statistics
$\mathbf{Z}_{N_{b-1}}$, $\mathbf{U}_{N_{b-1}}$ and
$\tilde{\boldsymbol{\beta}}_b^{(k-1)}$. Then the proximal gradient update is then performed. After the inner iteration converges, the statistics
$\mathbf{Z}_{N_b}$ and $\mathbf{U}_{N_b}$ are updated using the online estimator
$\tilde{\boldsymbol{\beta}}_b$ and used as historical summaries for the next batch.

\begin{algorithm}[htbp]
    \caption{Local PGD Batch Updating Algorithm for Streaming Data}
    \label{alg:local_pgd}
     \SetKwData{Left}{left}\SetKwData{This}{this}\SetKwData{Up}{up}
    \SetKwFunction{Union}{Union}\SetKwFunction{FindCompress}{FindCompress}
    \SetKwInOut{Input}{Input}\SetKwInOut{Output}{Output}
    
    \Input{The $b$-th data batch $\mathcal{D}_b=(\mathbf{X}^{(b)}, \boldsymbol{Y}^{(b)})$, expectile level $\tau$, penalty parameter $\lambda_b$, historical cumulative statistics $\mathbf{Z}_{N_{b-1}}$ and $\mathbf{U}_{N_{b-1}}$, the renewable estimator from the previous batch $\tilde{\boldsymbol{\beta}}_{b-1}$, cumulative sample size $N_b$, maximum number of inner iterations $K=100$, and convergence tolerance $\varepsilon=10^{-5}$.}
    
    \Output{The updated estimator for the $b$-th batch $\tilde{\boldsymbol{\beta}}_b$, and the updated historical cumulative statistics $\mathbf{Z}_{N_b}$ and $\mathbf{U}_{N_b}$.}
    
    \textbf{PGD initialization:}\\
    Set the initial iterate as the renewable estimator from the previous batch:
    $\tilde{\boldsymbol{\beta}}_b^{(0)}=\tilde{\boldsymbol{\beta}}_{b-1}$.
    
    \For{$k=1$ \KwTo $K$}{
        \textbf{1) Compute current-batch weights:}
        \begin{align*}
        w_{b,i}^{(k)}
        =
        \left|
        \tau
        -
        I\left(
        y_{b,i}
        <
        \boldsymbol{x}_{b,i}^{\top}
        \tilde{\boldsymbol{\beta}}_b^{(k-1)}
        \right)
        \right|,
        \quad
        \mathbf{W}^{(b)}_{\tilde{\boldsymbol{\beta}}_b^{(k-1)}}
        =
        \operatorname{diag}
        \{w_{b,1}^{(k)},\dots,w_{b,n_b}^{(k)}\}.
        \end{align*}
        
        \textbf{2) Compute the cumulative statistics for the $k$-th iteration:}
        \begin{align*}
            \mathbf{Z}_{N_b}^{(k)}
            =
            \mathbf{Z}_{N_{b-1}}
            +
            \mathbf{X}^{(b)\top}
            \mathbf{W}^{(b)}_{\tilde{\boldsymbol{\beta}}_b^{(k-1)}}
            \mathbf{X}^{(b)},\quad
            \mathbf{U}_{N_b}^{(k)}
            =
            \mathbf{U}_{N_{b-1}}
            +
            \mathbf{X}^{(b)\top}
            \mathbf{W}^{(b)}_{\tilde{\boldsymbol{\beta}}_b^{(k-1)}}
            \boldsymbol{Y}^{(b)}.
        \end{align*}
        
        \textbf{3) Proximal gradient update:}
        
        Compute the approximate gradient
        \[
            \boldsymbol{g}_{N_b}^{(k)}
            =
            \frac{1}{N_b}
            \left(
            \mathbf{Z}_{N_b}^{(k)}
            \tilde{\boldsymbol{\beta}}_b^{(k-1)}
            -
            \mathbf{U}_{N_b}^{(k)}
            \right).
        \]
        Then update the parameter using the soft-thresholding operator:
        \begin{align*}
            \eta
            =
            \left\|
            \frac{1}{N_b}
            \mathbf Z_{N_b}^{(k)}
            \right\|_1^{-1},\quad
            \tilde{\boldsymbol{\beta}}_b^{(k)}
            =
            \operatorname{prox}_{\eta\lambda_b}
            \left(
            \tilde{\boldsymbol{\beta}}_b^{(k-1)}
            -
            \eta \boldsymbol{g}_{N_b}^{(k)}
            \right).
        \end{align*}
        
        \eIf{
        $\left\|
        \tilde{\boldsymbol{\beta}}_b^{(k)}
        -
        \tilde{\boldsymbol{\beta}}_b^{(k-1)}
        \right\|_{\infty}
        < \varepsilon$}{
        Stop the iteration\;
        }{
        Continue until the maximum number of iterations $K$ is reached\;
        }
    }
    
    \textbf{Update historical cumulative statistics:}\\
    Recompute the current-batch weight matrix
    $\mathbf{W}^{(b)}_{\tilde{\boldsymbol{\beta}}_b}$
    based on the final estimator $\tilde{\boldsymbol{\beta}}_b$, and update
    \begin{align*}
        \mathbf{Z}_{N_b}
        =
        \mathbf{Z}_{N_{b-1}}
        +
        \mathbf{X}^{(b)\top}
        \mathbf{W}^{(b)}_{\tilde{\boldsymbol{\beta}}_b}
        \mathbf{X}^{(b)},\quad
        \mathbf{U}_{N_b}
        =
        \mathbf{U}_{N_{b-1}}
        +
        \mathbf{X}^{(b)\top}
        \mathbf{W}^{(b)}_{\tilde{\boldsymbol{\beta}}_b}
        \boldsymbol{Y}^{(b)}.
    \end{align*}
    
\end{algorithm}

The above procedure provides a sparse parameter estimation method for high-dimensional expectile regression in streaming data settings. However, as in the offline setting, the Lasso penalty introduces shrinkage bias toward zero. Consequently, the sparse updated estimator
$\tilde{\boldsymbol{\beta}}_b$ obtained at batch $b$ is not directly suitable for statistical inference. A further debiasing correction is therefore required. Let the Hessian-type matrix of the surrogate loss at the sparse updated estimator
$\tilde{\boldsymbol{\beta}}_b$ be
$\widetilde{\boldsymbol{\Sigma}}_b
=
\nabla^2\mathcal Q_{N_b}^{\mathrm{s}}(\tilde{\boldsymbol{\beta}}_b)
=
\mathbf Z_{N_b}/N_b$,
and let
$\widetilde{\boldsymbol{\Theta}}_b$
denote its approximate inverse. Then the debiased estimator for streaming data is defined as
\begin{equation}\label{eq:debias-stream}
\tilde{\boldsymbol{\beta}}_\text{de}
=
\tilde{\boldsymbol{\beta}}_b
+
\widetilde{\boldsymbol{\Theta}}_{b}
\left[
\frac{
\mathbf{U}_{N_b}
-
\mathbf{Z}_{N_b}\tilde{\boldsymbol{\beta}}_b
}{N_b}
\right].
\end{equation}

\subsection{Adaptive penalty parameter estimation}\label{sec2-sub2}
Note that the for each batch arrives, the online updating procedure no only depend on the cumulative statistic but also depend on the value of Lasso penalty $\lambda$. However in streaming scenario selecting the penalty parameter by cross-validation based on the global data requires continuous storage of historical samples. Moreover, if the penalty parameter is re-selected after the arrival of each data batch, the computational cost will increase substantially, making it difficult to meet the real-time updating requirement of online learning. Therefore, in this study we proposed an adaptive decay strategy for updating the penalty parameter.

According to high-dimensional statistical theory, it suggests that the penalty parameter typically satisfy
$\lambda_t \asymp \sqrt{\log p / N_t}$. Adopt this theoretical result, we derive the initial penalty parameter $\lambda_{cv}^{(1)}$ by the classical offline 5-fold cross-validation based on the first data batch $\mathcal{D}_1$. Then the 
the constant factor $c$ in this theoretical rate are then obtains, namely
$c_{opt}=\lambda_{cv}^{(1)}/\sqrt{\log p / n_1}$. For subsequent streaming batches $t\in\{2,\dots,b\}$, the algorithm no longer performs cross-validation repeatedly. Instead, the penalty parameter is deterministically updated according to the cumulative sample size $N_t$:
\begin{equation*}
\lambda_t
=
c_{opt}
\sqrt{\frac{\log p}{N_t}}.
\end{equation*}
Compared with a fixed penalty parameter, it gradually weakens the shrinkage strength as cumulative information increases, thereby reducing the risk of over-shrinkage in later batches.

\section{Online Precision Matrix Estimation with BIC criterion}\label{sec:precision}
In section~\ref{sec2-sub1}, we introduce the overall framework of DeCOER. However, this framework still faces two practical computational challenges. First, in high-dimensional settings, the empirical covariance matrix is often singular. Moreover, historical raw data cannot be repeatedly accessed in streaming scenarios, making conventional debiasing methods based on the inversion of the full-sample covariance matrix difficult to apply directly. Second, selecting the penalty parameter by cross-validation over the global data requires storing historical samples and substantially increases computational cost, which is incompatible with the real-time updating requirement of online learning. Motivated by these challenges, in this study we further investigates precision matrix estimation and penalty parameter selection strategies for streaming data.

In high-dimensional setting estimating the precision matrix
$\boldsymbol{\Theta}_{\tilde{\boldsymbol{\beta}}_b}$
is crucial for constructing the debiasing correction term. A commonly used approach is nodewise regression. However, as mensioned before in streaming data settings, raw data from historical batches are no longer retained after being processed. As a result, it is infeasible to repeatedly perform $p$ nodewise regressions based on the full-sample design matrix. Motivated by this, this study proposes a streaming precision matrix estimation algorithm that depends only on the cumulative covariance matrix $\mathbf Z_{N_b}$, thereby enabling online recursive estimation of the precision matrix in high-dimensional settings without accessing historical raw data.

Ideally, if the full sample were available, the global nodewise regression objective function are defined as:
\begin{equation}\label{eq:node}
\hat{\boldsymbol{\phi}}_{b,j}
=
\mathop{\arg\min}_{\boldsymbol{\phi}\in\mathbb R^{p-1}}
\left\{
\frac{1}{2N_b}
\sum_{t=1}^{b}
\left(
\boldsymbol{x}^{(t)}_{j}
-
\mathbf X^{(t)}_{-j}\boldsymbol{\phi}
\right)^\top
\mathbf W^{(t)}_{\tilde{\boldsymbol{\beta}}_t}
\left(
\boldsymbol{x}^{(t)}_{j}
-
\mathbf X^{(t)}_{-j}\boldsymbol{\phi}
\right)
+
\lambda^\text{node}_j\|\boldsymbol{\phi}\|_1
\right\},
\end{equation}
where $\boldsymbol{x}^{(t)}_j$ denotes the $j$-th column of the design matrix $\mathbf X^{(t)}$ for the $t$-th data batch, and $\mathbf X^{(t)}_{-j}$ denotes the design matrix with the $j$-th column removed. The weight matrix $\mathbf W^{(t)}_{\tilde{\boldsymbol{\beta}}_t}$ is evaluated at the online updated estimator $\tilde{\boldsymbol{\beta}}_t$ obtained from the $t$-th batch. Expanding the quadratic term in Eq.~\eqref{eq:node} gives,
\begin{equation}
\label{eq:node2}
\begin{aligned}
&\frac{1}{2N_b}
\sum_{t=1}^{b}
\left(
\boldsymbol{x}^{(t)}_{j}
-
\mathbf X^{(t)}_{-j}\boldsymbol{\phi}
\right)^\top
\mathbf W^{(t)}_{\tilde{\boldsymbol{\beta}}_t}
\left(
\boldsymbol{x}^{(t)}_{j}
-
\mathbf X^{(t)}_{-j}\boldsymbol{\phi}
\right)\\
=&
\frac{1}{2N_b}
\sum_{t=1}^{b}
\boldsymbol{x}^{(t)\top}_{j}
\mathbf W^{(t)}_{\tilde{\boldsymbol{\beta}}_t}
\boldsymbol{x}^{(t)}_{j}
-
\frac{1}{N_b}
\sum_{t=1}^{b}
\boldsymbol{\phi}^{\top}
\mathbf X^{(t)\top}_{-j}
\mathbf W^{(t)}_{\tilde{\boldsymbol{\beta}}_t}
\boldsymbol{x}^{(t)}_{j} 
+
\frac{1}{2N_b}
\sum_{t=1}^{b}
\boldsymbol{\phi}^{\top}
\mathbf X^{(t)\top}_{-j}
\mathbf W^{(t)}_{\tilde{\boldsymbol{\beta}}_t}
\mathbf X^{(t)}_{-j}
\boldsymbol{\phi}.
\end{aligned}
\end{equation}

According to previous section, the definition of the online covariance matrix are
$\widetilde{\boldsymbol{\Sigma}}_{b}
=
N_b^{-1}
\sum_{t=1}^{b}
\mathbf X^{(t)\top}
\mathbf W^{(t)}_{\tilde{\boldsymbol{\beta}}_t}
\mathbf X^{(t)}$,
hence the online covariance matrix contains the second-order moment information required for nodewise regression. Here, we split this matrix it with respect to the $j$-th variable as
\begin{equation*}
\widetilde{\boldsymbol{\Sigma}}_{b}
=
\begin{pmatrix}
\widetilde{\sigma}_{b,jj}
&
\widetilde{\boldsymbol{\sigma}}_{b,j,-j}\\
\widetilde{\boldsymbol{\sigma}}_{b,-j,j}
&
\widetilde{\boldsymbol{\Sigma}}_{b,-j,-j}
\end{pmatrix}.
\end{equation*}
Specifically,
\begin{equation*} 
\begin{aligned}
&\widetilde{\sigma}_{b,jj}
=
\frac{1}{N_b}
\sum_{t=1}^{b}
\boldsymbol{x}^{(t)\top}_{j}
\mathbf W^{(t)}_{\tilde{\boldsymbol{\beta}}_t}
\boldsymbol{x}^{(t)}_{j},
\quad
\widetilde{\boldsymbol{\sigma}}_{b,-j,j}
=
\frac{1}{N_b}
\sum_{t=1}^{b}
\mathbf X^{(t)\top}_{-j}
\mathbf W^{(t)}_{\tilde{\boldsymbol{\beta}}_t}
\boldsymbol{x}^{(t)}_{j},
\quad \\
&\widetilde{\boldsymbol{\Sigma}}_{b,-j,-j}
=
\frac{1}{N_b}
\sum_{t=1}^{b}
\mathbf X^{(t)\top}_{-j}
\mathbf W^{(t)}_{\tilde{\boldsymbol{\beta}}_t}
\mathbf X^{(t)}_{-j}.
\end{aligned}
\end{equation*}
Hence, after omitting the first term in~\eqref{eq:node2} which are independent of $\boldsymbol{\phi}$, the nodewise regression objective function can be equivalently reformed as:
\begin{equation}\label{eq:node-stream}
\hat{\boldsymbol{\phi}}_{b,j}
=
\mathop{\arg\min}_{\boldsymbol{\phi}\in\mathbb R^{p-1}}
\left\{
\frac{1}{2}
\boldsymbol{\phi}^{\top}
\widetilde{\boldsymbol{\Sigma}}_{b,-j,-j}
\boldsymbol{\phi}
-
\boldsymbol{\phi}^{\top}
\widetilde{\boldsymbol{\sigma}}_{b,-j,j}
+
\lambda^\text{node}_j\|\boldsymbol{\phi}\|_1
\right\}.
\end{equation}
Through this transformation, the reformed loss function defined in Eq.~\eqref{eq:node-stream} depends only on the online covariance matrix estimator and its corresponding sub-matrices, and therefore does not require revisiting historical raw data.

For illustration, we first consider the two-dimensional case, where the online covariance matrix are:
\begin{equation*}
\widetilde{\boldsymbol{\Sigma}}_{b}
=
\begin{pmatrix}
\widetilde{\sigma}_{b,11} & \widetilde{\sigma}_{b,12}\\
\widetilde{\sigma}_{b,21} & \widetilde{\sigma}_{b,22}
\end{pmatrix},
\quad
\widetilde{\sigma}_{b,kl}
=
\frac{1}{N_b}
\sum_{t=1}^{b}
\boldsymbol{x}^{(t)\top}_{k}
\mathbf W^{(t)}_{\tilde{\boldsymbol{\beta}}_t}
\boldsymbol{x}^{(t)}_{l}.
\end{equation*}
For node $j=1$, the estimator is obtained by solving the following optimization problem:
\begin{equation*}
\hat{\phi}_{b,1}
=
\mathop{\arg\min}_{\phi}
\left\{
\frac{1}{2N_b}
\sum_{t=1}^{b}
(\boldsymbol{x}^{(t)}_1-\phi \boldsymbol{x}^{(t)}_2)^\top
\mathbf W^{(t)}_{\tilde{\boldsymbol{\beta}}_t}
(\boldsymbol{x}^{(t)}_1-\phi \boldsymbol{x}^{(t)}_2)
+
\lambda^\text{node}_1|\phi|
\right\}.
\end{equation*}
Expanding the quadratic term and omitting the constant term independent of $\phi$, the solution equivalent to,
\begin{equation*}
\hat{\phi}_{b,1}
=
\mathop{\arg\min}_{\phi}
\left\{
\frac{1}{2}\widetilde{\sigma}_{b,22}\phi^2
-
\widetilde{\sigma}_{b,21}\phi
+
\lambda^\text{node}_1|\phi|
\right\}.
\end{equation*}
Similarly, for node $j=2$,
\begin{equation*}
\hat{\phi}_{b,2}
=
\mathop{\arg\min}_{\phi}
\left\{
\frac{1}{2}\widetilde{\sigma}_{b,11}\phi^2
-
\widetilde{\sigma}_{b,12}\phi
+
\lambda^\text{node}_2|\phi|
\right\}.
\end{equation*}

After obtaining the two nodewise regression coefficients define the coefficient matrix as
\begin{equation*}
\widehat{\boldsymbol{\Phi}}_{b}
=
\begin{pmatrix}
1 & -\hat{\phi}_{b,1}\\
-\hat{\phi}_{b,2} & 1
\end{pmatrix}.
\end{equation*}
Furthermore, define the scaling terms of node $j$ as:
\begin{equation*}
\hat d_{b,j}^2
=\frac{1}{N_b}
\sum_{t=1}^{b}
\boldsymbol{x}_j^{(t)\top}
\mathbf W^{(t)}_{\tilde{\boldsymbol{\beta}}_t}
\left(\boldsymbol{x}_j^{(t)}-
\hat{\phi}_{b,j}\boldsymbol{x}_l^{(t)}
\right)
\end{equation*}
Omit the penalty term, the first-order conditions for the two nodewise regressions are given as
\begin{equation}\label{eq:firstorder}
\widetilde{\sigma}_{b,22}\hat{\phi}_{b,1}
=
\widetilde{\sigma}_{b,21},
\quad
\widetilde{\sigma}_{b,11}\hat{\phi}_{b,2}
=
\widetilde{\sigma}_{b,12}.
\end{equation}
Hence the scale term can defined as:
\begin{equation*}
\begin{aligned}
\hat d_{b,1}^2
&=
\frac{1}{N_b}
\sum_{t=1}^{b}
\boldsymbol{x}_1^{(t)\top}
\mathbf W^{(t)}_{\tilde{\boldsymbol{\beta}}_t}
\left(
\boldsymbol{x}_1^{(t)}
-
\hat{\phi}_{b,1}\boldsymbol{x}_2^{(t)}
\right)
=
\widetilde{\sigma}_{b,11}
-
\hat{\phi}_{b,1}\widetilde{\sigma}_{b,21},\\
\hat d_{b,2}^2
&=
\frac{1}{N_b}
\sum_{t=1}^{b}
\boldsymbol{x}_2^{(t)\top}
\mathbf W^{(t)}_{\tilde{\boldsymbol{\beta}}_t}
\left(
\boldsymbol{x}_2^{(t)}
-
\hat{\phi}_{b,2}\boldsymbol{x}_1^{(t)}
\right)
=
\widetilde{\sigma}_{b,22}
-
\hat{\phi}_{b,2}\widetilde{\sigma}_{b,12}.
\end{aligned} 
\end{equation*}
Let
$\widehat{\boldsymbol D}_{b}^2
=
\operatorname{diag}
(\hat d_{b,1}^2,\hat d_{b,2}^2)$.
Combining the first-order conditions in Eq.~\eqref{eq:firstorder} with the symmetry of
$\widetilde{\boldsymbol{\Sigma}}_{b}$, we have
\begin{equation*}
\widehat{\boldsymbol{\Phi}}_{b}
\widetilde{\boldsymbol{\Sigma}}_{b}
=
\begin{pmatrix}
\hat d_{b,1}^2 & 0\\
0 & \hat d_{b,2}^2
\end{pmatrix}
=
\widehat{\boldsymbol D}_{b}^2.
\end{equation*}
Thus, the estimator of the inverse covariance matrix, 
$\widetilde{\boldsymbol{\Theta}}_{\tilde{\boldsymbol{\beta}}_b}$, is defined as
\begin{equation}\label{eq:precision}
\widetilde{\boldsymbol{\Theta}}_{b}
=
\widehat{\boldsymbol D}_{b}^{-2}
\widehat{\boldsymbol{\Phi}}_{b}.
\end{equation}

More generally, in the high-dimensional $p$-dimensional case, for each node $j=1,\dots,p$, nodewise regression after omitting the constant term leads to the following optimize function:
\begin{equation*}
\hat{\boldsymbol{\phi}}_{b,j}
=
\mathop{\arg\min}_{\boldsymbol{\phi}\in\mathbb R^{p-1}}
\left\{
\frac{1}{2}
\boldsymbol{\phi}^{\top}
\widetilde{\boldsymbol{\Sigma}}_{b,-j,-j}
\boldsymbol{\phi}
-
\boldsymbol{\phi}^{\top}
\widetilde{\boldsymbol{\sigma}}_{b,-j,j}
+
\lambda^{\text{node}}_j\|\boldsymbol{\phi}\|_1
\right\}.
\end{equation*}
Here, $\hat{\boldsymbol{\phi}}_{b,j}$ is a $(p-1)$-dimensional vector representing the nodewise regression coefficients obtained by using all variables except the $j$-th variable to explain the $j$-th variable. The coefficient matrix is then constructed as
\begin{equation*}
\widehat{\boldsymbol{\Phi}}_{b}
=
\begin{pmatrix}
1 & -\hat\phi_{b,12} & \cdots & -\hat\phi_{b,1p}\\
-\hat\phi_{b,21} & 1 & \cdots & -\hat\phi_{b,2p}\\
\vdots & \vdots & \ddots & \vdots\\
-\hat\phi_{b,p1} & -\hat\phi_{b,p2} & \cdots & 1
\end{pmatrix},
\end{equation*}
where $\hat\phi_{b,jl}$ denotes the coefficient of the $l$-th variable in the $j$-th nodewise regression $(l\neq j)$.

Similarly, define the scaling term for the $j$-th node as
$\hat d_{b,j}^2
=
\widetilde{\sigma}_{b,jj}
-
\hat{\boldsymbol{\phi}}_{b,j}^{\top}
\widetilde{\boldsymbol{\sigma}}_{b,-j,j}$.
In practical computation, to avoid numerical instability caused by excessively small scaling terms, set
\begin{equation*}
\hat d_{b,j}^2
=
\max
\left\{
\widetilde{\sigma}_{b,jj}
-
\hat{\boldsymbol{\phi}}_{b,j}^{\top}
\widetilde{\boldsymbol{\sigma}}_{b,-j,j},
c_0
\right\},
\end{equation*}
where $c_0>0$ is a constant, and in this study we sets $c_0=10^{-4}$. Define the diagonal matrix
$\widehat{\boldsymbol D}_{b}^2
=
\operatorname{diag}
(\hat d_{b,1}^2,\hat d_{b,2}^2,\ldots,\hat d_{b,p}^2)$.
Then the precision matrix can be computed based on Eq.~\eqref{eq:precision}, namely
$\widetilde{\boldsymbol{\Theta}}_{b}
=
\widehat{\boldsymbol D}_{b}^{-2}
\hat{\boldsymbol{\Phi}}_{b}$.

Finally, since cross-validation based on the global data cannot be directly implemented in streaming scenarios. We propose a BIC-based method for selecting the penalty parameter. For a given candidate parameter $\lambda^{\text{cand}}\in\Lambda$, define the residual sum of squares for the $j$-th nodewise regression as
\begin{equation*}
\text{RSS}_{b,j}(\lambda^{\text{cand}})
=
\frac{1}{N_b}
\sum_{t=1}^{b}
\left(
\boldsymbol{x}^{(t)}_{j}
-
\mathbf X^{(t)}_{-j}
\hat{\boldsymbol{\phi}}_{b,j}(\lambda^{\text{cand}})
\right)^\top
\mathbf W^{(t)}_{\tilde{\boldsymbol{\beta}}_t}
\left(
\boldsymbol{x}^{(t)}_{j}
-
\mathbf X^{(t)}_{-j}
\hat{\boldsymbol{\phi}}_{b,j}(\lambda^{\text{cand}})
\right).
\end{equation*}
Using the cumulative historical statistic $\mathbf Z_{N_b}$, this can be equivalently written as
\begin{equation*}
\text{RSS}_{b,j}(\lambda^{\text{cand}})
=
\widetilde{\sigma}_{b,jj}
-
2\hat{\boldsymbol{\phi}}_{b,j}(\lambda^{\text{cand}})^\top
\widetilde{\boldsymbol{\sigma}}_{b,-j,j}
+
\hat{\boldsymbol{\phi}}_{b,j}(\lambda^{\text{cand}})^\top
\widetilde{\boldsymbol{\Sigma}}_{b,-j,-j}
\hat{\boldsymbol{\phi}}_{b,j}(\lambda^{\text{cand}}).
\end{equation*}
Accordingly, the BIC criterion is defined as
\begin{equation*}
\text{BIC}_{b,j}(\lambda)
=
N_b\log\{\text{RSS}_{b,j}(\lambda)\}
+
\log(N_b)\cdot
\left\|
\hat{\boldsymbol{\phi}}_{b,j}(\lambda)
\right\|_0.
\end{equation*}
Finally, the optimal penalty parameter is selected as
\begin{equation*}
{\lambda_{b,j}^{*}}^\text{node}
=
\mathop{\arg\min}_{\lambda^{\text{cand}}\in\Lambda}
\text{BIC}_{b,j}(\lambda^{\text{cand}}).
\end{equation*}
Note that, the proposed online framework relies only on the online covariance estimator and does not require repeated access to historical raw data. At the same time, by introducing the BIC criterion, it avoids the additional computational burden caused by multi-grid cross-validation.

\begin{figure}[htbp]
    \centering
    \includegraphics[width=1\linewidth]{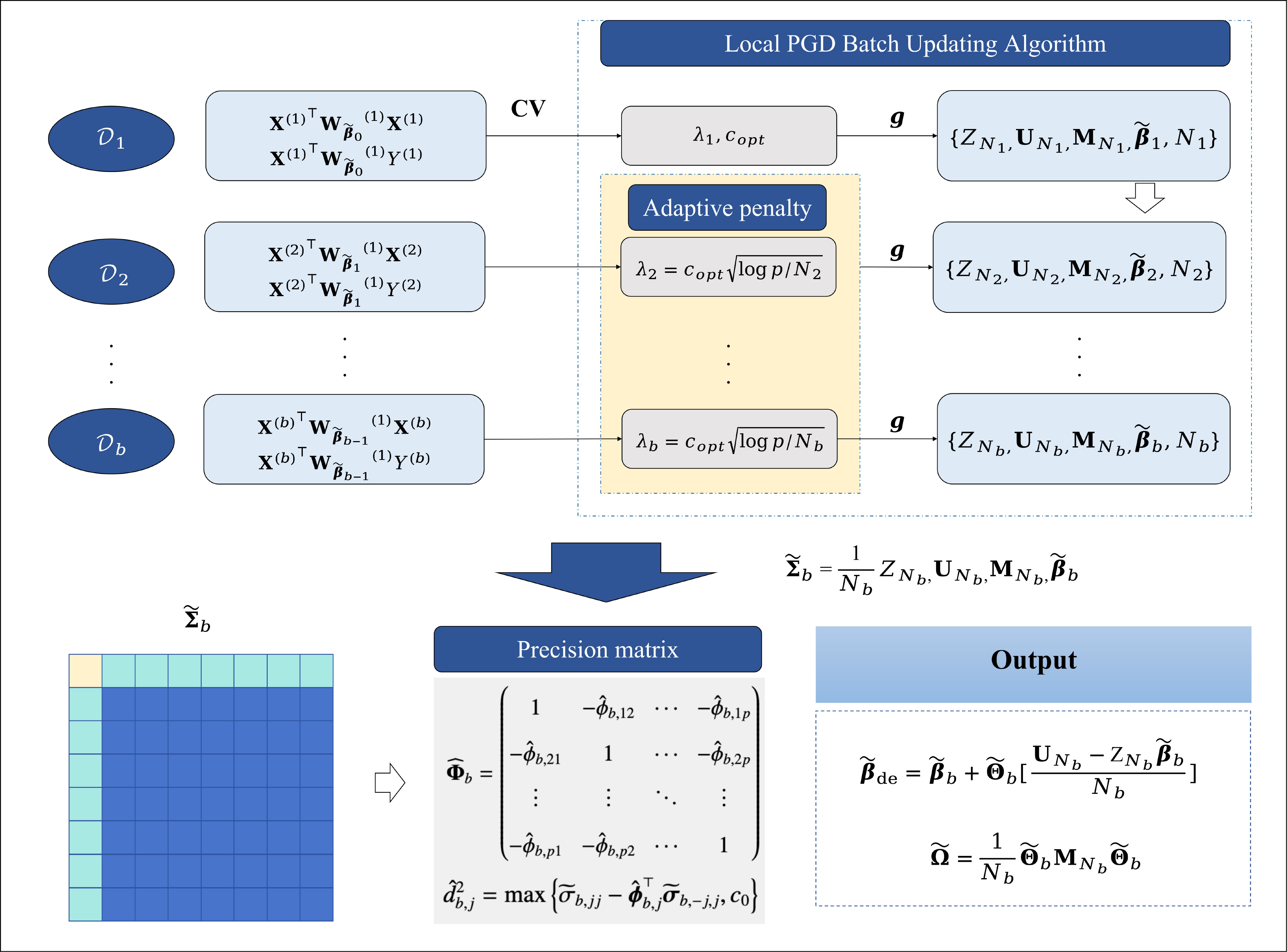}
    \caption{Flowchart for DeCOER}
    \label{fig:DeCOER}
\end{figure}

Figure~\ref{fig:DeCOER} illustrates the debiased renewable estimation framework. As shown in figure~\ref{fig:DeCOER}, after each data batch arrives, the algorithm first computes the recursive statistics based on the current batch data and the parameter estimate obtained from the previous batch. It then updates the sparse parameter estimate through the PGD algorithm (see algorithm~\ref{alg:local_pgd}) with a batch-specific penalty parameter. After the batch has been processed, the algorithm stores only the cumulative sufficient statistics rather than the historical raw data, thereby avoiding repeated storage and access to historical samples in streaming data settings. In addition, to reduce the computational cost in the parameter updating stage, DeCOER performs CV only on the first data batch to obtain the initial empirical optimal penalty parameter and calibrate the theoretical constant. For subsequent batches, the penalty parameter is adaptively decayed according to the cumulative sample size and the dimensionality.
In the debiasing stage, DeCOER constructs the precision matrix estimator based on the Hessian matrix
$\widetilde{\boldsymbol{\Sigma}}$.
Specifically, by partitioning
$\widetilde{\boldsymbol{\Sigma}}$,
the algorithm rewrites the traditional nodewise Lasso as an equivalent optimization problem that depends only on the cumulative covariance matrix. It then adopts the BIC criterion to select the penalty parameter in each nodewise regression, thereby obtaining a sparse pseudo-inverse estimator of the precision matrix.

\section{Large Sample Properties and Statistic Inference}\label{sec:theory}

In this section we establish the large-sample theoretical properties of the DeCOER method under a high-dimensional asymptotic framework with total sample size $N_b\to\infty$. Here, allow $p=p_{N_b}$, $s=s_{N_b}$ and the row sparsity of the precision matrix to vary with the cumulative sample size. Note that the estimator defined in Section~\ref{sec:Method} is the output of the PGD algorithm after a finite number of iterations, hence is not exactly equal to the ideal minimizer of the surrogate loss function. For theoretical analysis, define
$\ddot{\boldsymbol{\beta}}_b
=\mathop{\arg\min}_{\boldsymbol{\beta}\in\mathbb R^p}\mathcal Q_{N_b,\lambda}^{\mathrm{s}}(\boldsymbol{\beta})$
as the minimizer of the surrogate objective function. Let
$\ddot{\boldsymbol{\Sigma}}_b$ and $\ddot{\boldsymbol{\Theta}}_b$ denote the corresponding Hessian-type matrix and precision matrix estimator at this point.

We further introduce the following notation. Let $\lambda_{\min}(A)$ and $\lambda_{\max}(A)$ denote the minimum and maximum eigenvalues of a square matrix $A$, respectively. Let $\|\cdot\|_2$ denote the vector $\ell_2$-norm or the matrix spectral norm, and $\|\cdot\|_{\max}$ denote the elementwise maximum norm of a matrix. Let $S=\{j:\beta_{0j}\ne0\}$ denote the true support set with $s=|S|$. We assume that the following regularity conditions hold:

\begin{enumerate}[label=(C\arabic*),topsep=0pt]
\item \label{con:chap5-1}
The data batches are mutually independent. Let
$\epsilon_{ti}=y_{ti}-\boldsymbol{x}_{ti}^{\top}\boldsymbol{\beta}_0$
be the model error. Conditional on $\mathbf X=\boldsymbol{x}$, $\epsilon_{ti}$ are independent and identically distributed, and satisfy
$e_{\tau}(\epsilon_{ti}\mid \mathbf X=\boldsymbol{x})=0$
and
$\mathbb E(\epsilon_{ti}^2\mid \mathbf X=\boldsymbol{x})\le c_1<\infty$.

\item \label{con:chap5-2}
Conditional on $\mathbf X=\boldsymbol{x}$, the conditional density function
$f_{\epsilon_{ti}\mid \mathbf{X}}(\cdot)$ of $\epsilon_{ti}$ is continuous and bounded in a neighborhood of zero. That is, there exist constants $c_2>0$ and $\delta>0$ such that, for all $|u|\le\delta$,
$0<f_{\epsilon_{ti}\mid X}(u)\le c_2<\infty$.

\item \label{con:chap5-3}
The true parameter $\boldsymbol{\beta}_0$ is $s$-sparse, namely
$\|\boldsymbol{\beta}_0\|_0=s$. In addition, there exists a constant $c_3>0$ such that
$\|\boldsymbol{x}_{ti}\|_\infty\le c_3$.

\item \label{con:chap5-4}
Define the Hessian-type matrix at the true parameter as
$\boldsymbol{\Sigma}
=\mathbb E[
w_{\boldsymbol{\beta}_0,ti}\boldsymbol x_{ti}\boldsymbol x_{ti}^{\top}]$.
Assume that $\boldsymbol{\Sigma}$ is invertible, and define the population precision matrix
$\boldsymbol{\Theta}=\boldsymbol{\Sigma}^{-1}$, with $\boldsymbol{\Theta}_j$ denoting its $j$-th row. Further assume that there exists a constant $c_4>0$ such that, for any vector $\boldsymbol v$ satisfying
$\|\boldsymbol v_{S^c}\|_1\le3\|\boldsymbol v_S\|_1$, we have
$\boldsymbol v^\top\boldsymbol{\Sigma}\boldsymbol v\ge c_4\|\boldsymbol v\|_2^2$.

\item \label{con:chap5-5}
The online Lasso penalty parameter satisfies
$\lambda_b\asymp\sqrt{\log p/N_b}$ and $s\lambda_b\to0$.

\item \label{con:chap5-6}
Let $\boldsymbol{\phi}_{0,j}$ denote the population nodewise regression coefficient of the $j$-th variable on the remaining variables, and define
$s_{\Theta,j}=1+\|\boldsymbol{\phi}_{0,j}\|_0$.
For any fixed coordinate $j$, assume that
$\|\boldsymbol{\phi}_{0,j}\|_1=O(1)$,
$d_{0,j}^2\ge\kappa_0>c_0$, and that
$\boldsymbol{\Sigma}_{-j,-j}$ satisfies a restricted eigenvalue condition along nodewise sparse directions.

\item \label{con:chap5-7}
The nodewise regression penalty parameters selected by BIC have the same asymptotic order, denoted by $\lambda_b^{\mathrm{node}}$, and satisfy
$\lambda_b^{\mathrm{node}}
\asymp
\sqrt{\log p/N_b}+s\lambda_b$,
$\lambda_b^{\mathrm{node}}=o(1)$,
$s_{\Theta,j}\lambda_b^{\mathrm{node}}\to0$, and
$s_{\Theta,j}\lambda_b^{\mathrm{node}}\sqrt{\log p}\to0$.

\item \label{con:chap5-8}
The gradient of the surrogate loss at the true parameter satisfies
$\mathbb P\{
\|\nabla\mathcal Q_b^{\mathrm{s}}(\boldsymbol{\beta}_0)\|_\infty
\le
\lambda_b/2
\}\rightarrow 1$.
Moreover, there exists a constant $\kappa>0$ such that, for any
$\boldsymbol v\in\mathcal C(S,3)$, the following inequality holds with probability tending to one:
\begin{equation*}
\mathcal Q_b^{\mathrm{s}}
(\boldsymbol{\beta}_0+\boldsymbol v)
-
\mathcal Q_b^{\mathrm{s}}
(\boldsymbol{\beta}_0)
-
\nabla\mathcal Q_b^{\mathrm{s}}
(\boldsymbol{\beta}_0)^\top\boldsymbol v
\ge
\kappa\|\boldsymbol v\|_2^2.
\end{equation*}

\item \label{con:chap5-9}
The sparsity level and penalty parameters satisfy
$\sqrt{N_b}s^2\lambda_b^2\to0$ and
$\sqrt{N_b}\lambda_b^{\mathrm{node}}s\lambda_b\to0$.

\item \label{con:chap5-10}
For any fixed coordinate $j$, the random variable
$\boldsymbol{\Theta}_j w_{\boldsymbol{\beta}_0,ti}\epsilon_{ti}\boldsymbol x_{ti}$
satisfies the Lindeberg condition, and
$0<\Omega_{jj}<\infty$, where
$\boldsymbol{\Omega}
=
\boldsymbol{\Theta}
\mathbb E[
w_{\boldsymbol{\beta}_0,ti}^2\epsilon_{ti}^2
\boldsymbol x_{ti}\boldsymbol x_{ti}^{\top}]
\boldsymbol{\Theta}^{\top}$.

\item \label{con:chap5-11}
Let $\mathcal F_b$ denote the one-step proximal gradient mapping. Assume that there exists a constant $0<q<1$, independent of $b$, such that, with probability tending to one,
$\|\mathcal F_b(\boldsymbol{\beta})-\ddot{\boldsymbol{\beta}}_b\|_1
\le q\|\boldsymbol{\beta}-\ddot{\boldsymbol{\beta}}_b\|_1$
for any $\boldsymbol{\beta}$ in a neighborhood of $\ddot{\boldsymbol{\beta}}_b$.

\item \label{con:chap5-12}
Let $K_b$ denote the number of PGD iterations actually performed for the $b$-th batch, and define
$d_{b,0}=\|\tilde{\boldsymbol{\beta}}_{b-1}-\ddot{\boldsymbol{\beta}}_b\|_1$ and
$\delta_b=q^{K_b}d_{b,0}$. Assume that the number of iterations $K_b$ or the stopping tolerance $\varepsilon$ satisfies
\begin{equation*}
\delta_b=o_p(s\lambda_b),
\quad
\sqrt{N_b}
\left\{
\lambda_b^{\mathrm{node}}\delta_b+\delta_b^2
+\lambda_b
\max_j
\left\|
\widetilde{\boldsymbol{\Theta}}_{b,j}
-
\ddot{\boldsymbol{\Theta}}_{b,j}
\right\|_1
\right\}
=o_p(1).
\end{equation*}

\item \label{con:chap5-13}
The perturbation induced by frozen historical weights based on the actual PGD output satisfies
\begin{equation*}
\left\|
\frac1{N_b}\sum_{t=1}^{b-1}
\mathbf X^{(t)\top}
\left(
\mathbf W_{\tilde{\boldsymbol{\beta}}_t}^{(t)}
-\mathbf W_{\boldsymbol{\beta}_0}^{(t)}
\right)\boldsymbol\epsilon^{(t)}
\right\|_\infty
=o_p(N_b^{-1/2}).
\end{equation*}
\end{enumerate}

Conditions~\ref{con:chap5-1}--\ref{con:chap5-2} are standard assumptions for ER. Conditions~\ref{con:chap5-3}--\ref{con:chap5-5} specify the sparsity of the true parameter, boundedness of covariates, the restricted eigenvalue condition for the Hessian-type matrix, and the asymptotic order of the Lasso penalty parameter. Conditions~\ref{con:chap5-6}--\ref{con:chap5-7} are imposed for nodewise regression and precision matrix estimation. They require the population nodewise regression coefficients to be sparse and specify the asymptotic order of the nodewise penalty parameter. Condition~\ref{con:chap5-8} controls the gradient size and local curvature of the surrogate loss at the true parameter, ensuring identifiability of the surrogate objective along sparse directions. Condition~\ref{con:chap5-9} further restricts the growth rates of sparsity and penalty parameters. Condition~\ref{con:chap5-10} guarantees that the leading stochastic term satisfies a central limit theorem for fixed coordinates. Condition~\ref{con:chap5-11} is a local contraction condition for the finite-step PGD iteration, which typically holds when the surrogate objective satisfies restricted strong convexity, the step size is properly chosen, and the local support set is relatively stable. Condition~\ref{con:chap5-12} further requires that the finite-iteration error and its induced perturbation in precision matrix estimation do not affect the first-order asymptotic distribution at the $\sqrt{N_b}$ scale. In practice, this condition can be satisfied by increasing the number of iterations or by using a stopping tolerance that varies with the cumulative sample size. Condition~\ref{con:chap5-13} controls the overall surrogate error induced by using frozen historical weights instead of the true weights, ensuring that the historical-weight approximation does not alter the leading stochastic term. Based on the above regularity conditions, we first present the following lemmas.

\begin{lemma}\label{DeCOER-L1}
Under Conditions~\ref{con:chap5-1}--\ref{con:chap5-5} and Condition~\ref{con:chap5-8}, the minimizer of the surrogate objective satisfies
\begin{equation*}
\|\ddot{\boldsymbol{\beta}}_b-\boldsymbol{\beta}_0\|_2=O_p(\sqrt{s}\lambda_b),
\qquad
\|\ddot{\boldsymbol{\beta}}_b-\boldsymbol{\beta}_0\|_1=O_p(s\lambda_b).
\end{equation*}
If $s\lambda_b\to0$, then
$\|\ddot{\boldsymbol{\beta}}_b-\boldsymbol{\beta}_0\|_1=o_p(1)$.
\end{lemma}

\begin{lemma}\label{DeCOER-L2}
Under Conditions~\ref{con:chap5-1}--\ref{con:chap5-3}, for any
$\boldsymbol{\beta}_1$ and $\boldsymbol{\beta}_2$ in a neighborhood of the true parameter, let
$\boldsymbol{\Delta}=\boldsymbol{\beta}_1-\boldsymbol{\beta}_2$, and define the first-order expansion remainder of the surrogate gradient as:
$$\boldsymbol r_{g,b}(\boldsymbol{\beta}_1,\boldsymbol{\beta}_2)
:=
\nabla\mathcal Q_b^{\mathrm{s}}(\boldsymbol{\beta}_1)
-\nabla\mathcal Q_b^{\mathrm{s}}(\boldsymbol{\beta}_2)
-\boldsymbol{\Sigma}_b(\boldsymbol{\beta}_2)\boldsymbol{\Delta},$$
which satisfied
$\|\boldsymbol r_{g,b}(\boldsymbol{\beta}_1,\boldsymbol{\beta}_2)\|_\infty
=O_p(\|\boldsymbol{\Delta}\|_1^2)$.
\end{lemma}

\begin{lemma}\label{DeCOER-L3}
Under Conditions~\ref{con:chap5-6}--\ref{con:chap5-8}, for
$(\boldsymbol{\Sigma}_b^*,\boldsymbol{\Theta}_b^*)\in
\{(\ddot{\boldsymbol{\Sigma}}_b,\ddot{\boldsymbol{\Theta}}_b),
(\widetilde{\boldsymbol{\Sigma}}_b,\widetilde{\boldsymbol{\Theta}}_b)\}$,
the precision matrix estimator obtained from nodewise regression satisfies:
$$\|
\mathbf I-\boldsymbol{\Theta}_b^*\boldsymbol{\Sigma}_b^*
\|_{\max}
=O_p(\lambda_b^{\mathrm{node}}),\quad\max_j\|\boldsymbol{\Theta}_{b,j}^*\|_1=O_p(1).$$
Furthermore, for any fixed coordinate $j$,
$\|\boldsymbol{\Theta}_{b,j}^*-\boldsymbol{\Theta}_j\|_1
=O_p(s_{\Theta,j}\lambda_b^{\mathrm{node}})=o_p(1)$.
\end{lemma}

Lemma~\ref{DeCOER-L1} establishes the basic convergence rates of the ideal minimizer of the surrogate objective, which are used to control the initial sparse estimation error. Lemma~\ref{DeCOER-L2} characterizes the nonlinear remainder in the surrogate gradient expansion between two parameter points. Lemma~\ref{DeCOER-L3} establishes the approximate inverse property and row-wise convergence of the online nodewise precision matrix estimator.

To further relate the debiased estimator constructed from the ideal surrogate minimizer to that constructed from the actual finite-step PGD output, define the auxiliary debiased estimator corresponding to the ideal surrogate minimizer as
\begin{equation}\label{eq:DeCOER-debias-ideal}
\ddot{\boldsymbol{\beta}}_{\mathrm{de}}
=
\ddot{\boldsymbol{\beta}}_b
-
\ddot{\boldsymbol{\Theta}}_b
\nabla\mathcal Q_{N_b}^{\mathrm{s}}(\ddot{\boldsymbol{\beta}}_b).
\end{equation}
The DeCOER estimator then has the following large-sample properties.

\begin{theorem}\label{DeCOER-T1}
Under Conditions~\ref{con:chap5-1}--\ref{con:chap5-9} and Condition~\ref{con:chap5-13}, the debiased estimator corresponding to the ideal surrogate minimizer satisfies
\begin{equation*}
\ddot{\boldsymbol{\beta}}_{\mathrm{de}}-\boldsymbol{\beta}_0
=
\ddot{\boldsymbol{\Theta}}_b
\frac1{N_b}\sum_{t=1}^b
\mathbf X^{(t)\top}
\mathbf W_{\boldsymbol{\beta}_0}^{(t)}\boldsymbol\epsilon^{(t)}
+
\boldsymbol r_b,
\quad
\|\boldsymbol r_b\|_\infty=o_p(N_b^{-1/2}).
\end{equation*}
\end{theorem}

\begin{theorem}\label{DeCOER-T2}
Under Theorem~\ref{DeCOER-T1} and Condition~\ref{con:chap5-10}, for any fixed coordinate $j$,
\begin{equation*}
\sqrt{N_b}
\left(
\ddot{\beta}_{\mathrm{de},j}-\beta_{0j}
\right)
\xrightarrow{d}
N(0,\Omega_{jj}).
\end{equation*}
\end{theorem}

\begin{theorem}\label{DeCOER-T3}
Under Theorem~\ref{DeCOER-T2} and Conditions~\ref{con:chap5-11}--\ref{con:chap5-13}, the finite-step PGD output satisfies
\begin{equation*}
\|\tilde{\boldsymbol{\beta}}_b-\ddot{\boldsymbol{\beta}}_b\|_1
\le q^{K_b}d_{b,0}=o_p(s\lambda_b)
\quad
\text{and}
\quad
\sqrt{N_b}
\|\tilde{\boldsymbol{\beta}}_{\mathrm{de}}
-\ddot{\boldsymbol{\beta}}_{\mathrm{de}}\|_\infty=o_p(1).
\end{equation*}
Therefore, for any fixed coordinate $j$,
\begin{equation*}
\sqrt{N_b}
\left(
\tilde\beta_{\mathrm{de},j}-\beta_{0j}
\right)
\xrightarrow{d}
N(0,\Omega_{jj}).
\end{equation*}
\end{theorem}

Theorem~\ref{DeCOER-T1} provides the first-order asymptotic linear representation of the debiased estimator based on the ideal surrogate minimizer and shows that the remainder term is negligible at the $N_b^{-1/2}$ scale. Theorem~\ref{DeCOER-T2} further establishes asymptotic normality for any fixed coordinate of the ideal debiased estimator. Finally, Theorem~\ref{DeCOER-T3} extends these results to the actual finite-step PGD output. It shows that, as long as the finite-iteration error and the perturbations induced by it are negligible at the $\sqrt{N_b}$ scale, the practically computed debiased estimator is asymptotically equivalent to the auxiliary debiased estimator. Therefore, DeCOER retains fixed-coordinate asymptotic normality in practical online computation for streaming data, providing the theoretical foundation for confidence interval construction and hypothesis testing.

Finally, based on the above theoretical properties, we present the online statistical inference framework of DeCOER in Algorithm~\ref{alg:online_debias_main}.
According to Algorithm~\ref{alg:online_debias_main}, first estimate sparse estimator with the local PGD iterative update algorithm (see Algorithm~\ref{alg:local_pgd}). Then estimate the score variance matrix, defined as
\begin{equation*}
\mathbf M_{N_b}
=
\mathbf M_{N_{b-1}}
+
\mathbf X^{(b)\top}
\operatorname{diag}
\left\{
\left(w_i^{(b)}(\tilde{\boldsymbol{\beta}}_b)\right)^2
\left(\tilde\epsilon_i^{(b)}\right)^2
\right\}_{i=1}^{n_b}
\mathbf X^{(b)},
\end{equation*}
where 
$\tilde\epsilon_i^{(b)}
=
y_i^{(b)}
-
\boldsymbol{x}_i^{(b)\top}
\tilde{\boldsymbol{\beta}}_b$ is the estimated error term. Next estimate precision matrix $\widetilde{\boldsymbol{\Theta}}_{b}$ and derive the debiased estimator. After obtaining the debiased estimator $\tilde{\boldsymbol{\beta}}_{\mathrm{de}}$ up to the $b$-th batch, the asymptotic covariance matrix is estimated with a sandwich form:
$\widetilde{\boldsymbol{\Omega}}_b
=
N_b^{-1}
\widetilde{\boldsymbol{\Theta}}_{b}
\mathbf M_{N_b}
\widetilde{\boldsymbol{\Theta}}_{b}^{\top}$.

\begin{algorithm}[htbp]
    \caption{Online Debiased Expectile Regression for High-Dimensional Streaming Data (DeCOER)}
    \label{alg:online_debias_main}
    \SetKwData{Left}{left}\SetKwData{This}{this}\SetKwData{Up}{up}
    \SetKwFunction{Union}{Union}\SetKwFunction{FindCompress}{FindCompress}
    \SetKwInOut{Input}{Input}\SetKwInOut{Output}{Output}
    
    \Input{Streaming observations $\mathcal{D}_{(b)}=\{(\mathbf{X}^{(t)}, Y^{(t)})\}_{t=1}^b$, expectile level $\tau$, number of local PGD iterations $K$, and candidate grid $\lambda^\mathrm{cand}$.}
    \Output{Debiased estimator $\tilde{\boldsymbol{\beta}}_{\mathrm{de}}$ at level $\tau$ and online covariance matrix $\widetilde{\boldsymbol{\Omega}}$.}
    
    \textbf{Initialization:}
    $\tilde{\boldsymbol{\beta}}_0=\boldsymbol{0}$, $\mathbf Z_0=\mathbf0$, $\mathbf U_0=\boldsymbol{0}$, $\mathbf M_0=\mathbf0$, and $N_0=0$\;
    
    \For{$t = 1$ \KwTo $b$}{
        Update the cumulative sample size $N_t = N_{t-1} + n_t$\;
        
        \eIf{$t == 1$}{
            Perform cross-validation to obtain the penalty parameter and theoretical scaling constant:
            $\lambda_{cv}^{(1)},c_{opt} = \lambda_{cv}^{(1)} / \sqrt{\log p / n_1}$\;
        }{
            Update the decaying penalty parameter according to the asymptotic theory:
            $\lambda_t = c_{opt} \sqrt{\log p / N_t}$\;
        }
        
        Perform \textbf{Algorithm~\ref{alg:local_pgd}}  and obtain
        $\{\tilde{\boldsymbol{\beta}}_t, \mathbf{Z}_{N_t}, \mathbf{U}_{N_t}\}.
        $.

        Compute the residuals and weights, and update the score variance matrix:
        \[
        \mathbf{M}_{N_t}
        =
        \mathbf{M}_{N_{t-1}}
        +
        \mathbf{X}^{(t)\top}
        \operatorname{diag}
        \left\{
        \left(w_i^{(b)}(\tilde{\boldsymbol{\beta}}_b)\right)^2
\left(\tilde\epsilon_i^{(b)}\right)^2
        \right\}
        \mathbf{X}^{(t)}.
        \]
        
        Store the summary information up to data batch $t$ and release the local data $\mathcal{D}_t$\;
    }
    
    \textbf{Step 2: Online line debiased estimation:}
    
    Set $\widetilde{\boldsymbol{\Sigma}}_b = \mathbf{Z}_{N_b} / N_b$, perform online BIC-based penalty selection, and obtain the precision matrix estimator $\widetilde{\boldsymbol{\Theta}}_{b}$\;

    Compute the debiased estimator
    \[
    \tilde{\boldsymbol{\beta}}_{de}
    =
    \tilde{\boldsymbol{\beta}}_b
    +
    \widetilde{\boldsymbol{\Theta}}_{b}
    \left[
    \frac{\mathbf{U}_{N_b}-\mathbf{Z}_{N_b} \tilde{\boldsymbol{\beta}}_b}{N_b}
    \right].
    \]
    
    \textbf{Step 3: Online asymptotic covariance estimation:}
    \[
    \widetilde{\boldsymbol{\Omega}}
    =
    \frac{1}{N_b}
    \widetilde{\boldsymbol{\Theta}}_{b}
    \mathbf{M}_{N_b}
    \widetilde{\boldsymbol{\Theta}}_{b}^{\top}.
    \]
\end{algorithm}

By Theorem~\ref{DeCOER-T3}, for any fixed coordinate $j$,
\begin{equation*}
\frac{\sqrt{N_b}
\left(
\tilde\beta_{\mathrm{de},j}-\beta_{0j}
\right)}
{\sqrt{\widetilde\Omega_{b,jj}}}
\xrightarrow{d}
N(0,1).
\end{equation*}
Therefore, the standard error of the $j$-th regression coefficient can be estimated by
\begin{equation}\label{eq:DeCOER-se}
\widetilde{\operatorname{se}}
\left(
\tilde{\beta}_{de,j}
\right)
=
\sqrt{\frac{\widetilde{\Omega}_{b,jj}}{N_b}}.
\end{equation}
This standard error can be used to construct confidence intervals and conduct significance tests for any regression coefficient. For example, for the $j$-th coefficient, consider the null hypothesis
$H_0:\beta_j=0$. Based on Theorem~\ref{DeCOER-T3}, the large-sample $z$ statistic is
$
z_j
=
\tilde{\beta}_{\mathrm{de},j}
/
\widetilde{\operatorname{se}}
(
\tilde{\beta}_{\mathrm{de},j}
)$.
Under the null hypothesis,
$z_j \xrightarrow{d}N(0,1)$.
Thus, for a given significance level $\alpha$, the null hypothesis $H_0$ is rejected when
$|z_j|>z_{1-\alpha/2}$, indicating that the $j$-th regression coefficient is significantly different from zero; otherwise, the null hypothesis cannot be rejected.

It should be noted that, to improve computational efficiency, this study performs the global debiasing correction only once after all data up to the $b$-th batch have arrived. Therefore, during the online updating process, the un-debiased estimator from the previous step is used as the initial value for the next iteration. Nevertheless, the numerical simulations show that this approximation has only a minor effect on parameter estimation. Even when the debiasing correction is performed only once after the final batch arrives, the proposed method still achieves favorable asymptotic properties and convergence performance.

\section{Numerical simulation}
\label{sec:simulation}
\subsection{Model Setting and Evaluation Criteria}\label{sec6-sec1}

Assume that the data are generated from model~\eqref{eq:DeCOER-sim}:
\begin{equation}\label{eq:DeCOER-sim}
y_{ti} = \boldsymbol{x}_{ti}^\top \boldsymbol{\beta}^\star + (\boldsymbol{x}_{ti}^\top \boldsymbol{\gamma}) \cdot \epsilon_{ti},
\end{equation}
where the covariate vector is defined as
$\boldsymbol{x}_{ti} = \left(1, \sigma(\mathbf{Z}_{ti})^\top\right)^\top \in \mathbb{R}^{p+1}$ and
$\mathbf{Z}_{ti} \sim N(\boldsymbol{0}, \boldsymbol{I}_p)$ are i.i.d distribution. Here, $\sigma(\cdot)$ denotes a transformation function ensure the sign and scale of the expectile function. Denote $\epsilon_{ti}$ as error term and in this study, we consider two cases: (1) the standard normal distribution $\epsilon_{ti}\sim N(0,1)$ and (2) the heavy-tailed distribution $\epsilon_{ti}\sim t(3)$.

For the true parameters, we set dimension $p=200$ and the number of true nonzero signals is $s=10$. And the baseline regression coefficient vector is specified as
$\boldsymbol{\beta}^\star = (1, \underbrace{2, 2, \dots, 2}_{s}, 0, \dots, 0)^\top$.
In addition, we consider two variance settings:
\begin{enumerate}[label=(\arabic*),topsep=0pt]
    \item Homoscedastic setting:
    $\boldsymbol{\gamma} = (1, 0, \dots, 0)^\top$; the transformation function is the identity transformation, namely
    $\sigma(\mathbf{Z}_{ti}) = \{z_{ti,1}, \dots, z_{ti,p}\}^\top$.
    \item Heteroscedastic setting:
    $\boldsymbol{\gamma} = (1, 1, 0.5, 0, \ldots, 0)^\top$; the transformation function is
    \begin{equation*}
    \sigma(\mathbf{Z}_{ti}) = \{ \underbrace{\Phi(z_{ti,1}), \Phi(z_{ti,2})}_{s}, \underbrace{0, \dots, 0}_{p-2} \}^\top,
    \end{equation*}
\end{enumerate}
where $\Phi(\cdot)$ is the cumulative distribution function of the standard normal distribution, which maps the values into the interval $(0,1)$.

To examine the performance of the proposed method under different streaming data scenarios, we consider the following data-generating settings:
(1) the total sample size is fixed at $N_b=100,000$, while the batch size varies over
$n_t\in\{100,200,500,1000\}$;
(2) the batch size is fixed at $n_t=200$, and as
$b\in\{10,20,50,100\}$ increases, the total sample size $N_b$ increases from $2000$ to $20000$.

For comparison, we consider the following three benchmark methods.
(1) Oracle, which assumes that the true set of nonzero variables is known and performs standard ER with full data using only the true signal variables. 
(2) DeOffline, which performs debiased ER using the full static data and serves as the benchmark optimal method in the high-dimensional setting.
(3) LaOnline, an online high-dimensional ER method that performs sparse estimation based only on streaming data without debiasing.

We considerate the following metrics to evaluate method performance:
\begin{itemize}
    \item Mean squared error (MSE): 
    $$\text{MSE}(\boldsymbol{\beta}) = \frac{1}{M} \sum_{m=1}^{M} \left( \frac{1}{s} \sum_{j \in \mathcal{S}} (\tilde{\beta}_j^{(m)} - \beta_j^\star)^2 \right).$$

    \item Standard deviation (SD) and empirical standard error (SE): 
   $$ \text{SD} = \frac{1}{s} \sum_{j \in \mathcal{S}} \sqrt{\frac{1}{M-1} \sum_{m=1}^{M} \left(\tilde{\beta}_j^{(m)} - \bar{\beta}_j\right)^2}, \quad
    \text{SE} = \frac{1}{M \cdot s} \sum_{m=1}^{M} \sum_{j \in \mathcal{S}} \widetilde{\text{se}}_j^{(m)},$$
   
    where $\widetilde{\text{se}}_j^{(m)}$ is computed according to Eq.~\eqref{eq:DeCOER-se}.

    \item Confidence interval coverage probability (CP): 
    $$\text{CP} = \frac{1}{M \cdot s} \sum_{m=1}^{M} \sum_{j \in \mathcal{S}} 
    \boldsymbol{I} \left( \beta_j^\star \in \left[\tilde{\beta}_j^{(m)} \pm z_{1-\alpha/2} \cdot \widetilde{\text{se}}_j^{(m)}\right] \right).$$
    In this study we set the significant level $\alpha = 0.05$
\end{itemize}

In addition, we also consider  False positive rate (FPR, or type-I error) for evaluate the performance of theoretical results. Here, for the LaOnline method, it is defined as the proportion of truly zero parameters that are incorrectly selected as nonzero; for debiased methods such as DeOffline and DeCOER, it is computed as the type-I error rate at the significance level $\alpha=0.05$. For each model setting, we conduct independent simulation $M=200$.

\subsection{Main Simulation Results}\label{sec6-sub2}

Tables~\ref{tab:DeCOER-homo-fix} and~\ref{tab:DeCOER-heter-fix} report the simulation results of the proposed method and the three comparison methods under the fixed total sample size setting $N_b=100,000$ when $\tau=0.25$. The main findings are as follows:
\begin{enumerate}[label=(\arabic*)]
    \item When the total sample size $N_b$ is fixed, the MSE of Oracle and DeOffline remain essentially unchanged as the local batch size $n_t$ varies. By contrast, the proposed DeCOER method also shows highly stable MSE values under different $n_t$ settings, with magnitudes almost identical to Oracle, indicating that DeCOER can effectively approximate the optimal estimation performance in online settings.

    \item Although the MSE of LaOnline decreases as the block size $n_t$ increases, its overall accuracy remains substantially worse than that of DeCOER. Meanwhile, the SD values of DeCOER and LaOnline are very close, indicating that the performance improvement of DeCOER mainly comes from a substantial reduction in estimation bias.

    \item In terms of uncertainty estimation, the SD and SE of DeCOER are highly consistent under all settings, and its confidence interval coverage remains stable around the nominal level 0.95.  In addition, the FPR of DeCOER remains stable around the theoretical level, whereas the FPR of LaOnline exhibits substantial fluctuations. This further demonstrates that introducing a debiasing procedure is crucial for ensuring valid and statistical inference in streaming data settings.
\end{enumerate}

\begin{table}[htbp]
\footnotesize
   \centering
  \renewcommand{\arraystretch}{0.6}
  \caption{Performance of Homogeneous model, with fixed the whole sample size $N_b=100,000 (\tau=0.25)$}
    \begin{tabular}{ccccccccccc}
\toprule
Method & MSE & SD & SE & CP & FPR & MSE & SD & SE & CP & FPR \\
\midrule
\multicolumn{11}{c}{$\epsilon \sim N(0,1)$} \\
\midrule
      & \multicolumn{5}{c}{$n_t = 100$} & \multicolumn{5}{c}{$n_t = 200$} \\
\cmidrule(lr){2-6}\cmidrule(lr){7-11}
Oracle & 0.011 & 0.011 & 0.011 & 0.947 & $\diagup$ & 0.011 & 0.011 & 0.011 & 0.947 & $\diagup$ \\
DeOffline & 0.011 & 0.011 & 0.011 & 0.948 & 0.049 & 0.011 & 0.011 & 0.011 & 0.948 & 0.049 \\
LaOnline & 0.067 & 0.011 & 0.011 & 0.410 & 0.024 & 0.035 & 0.011 & 0.010 & 0.678 & 0.136 \\
DeCOER & 0.011 & 0.011 & 0.011 & 0.942 & 0.047 & 0.011 & 0.011 & 0.010 & 0.939 & 0.051 \\
\cmidrule(lr){2-6}\cmidrule(lr){7-11}
      & \multicolumn{5}{c}{$n_t = 500$} & \multicolumn{5}{c}{$n_t = 1000$} \\
\cmidrule(lr){2-6}\cmidrule(lr){7-11}
Oracle & 0.011 & 0.011 & 0.011 & 0.947 & $\diagup$& 0.011 & 0.011 & 0.011 & 0.947 & $\diagup$ \\
DeOffline & 0.011 & 0.011 & 0.011 & 0.948 & 0.049 & 0.011 & 0.011 & 0.011 & 0.948 & 0.049 \\
LaOnline & 0.031 & 0.011 & 0.010 & 0.741 & 0.241 & 0.038 & 0.011 & 0.010 & 0.648 & 0.122 \\
DeCOER & 0.011 & 0.011 & 0.010 & 0.937 & 0.053 & 0.011 & 0.011 & 0.010 & 0.938 & 0.051 \\

\midrule
\multicolumn{11}{c}{$\epsilon \sim t(3)$} \\
\midrule
      & \multicolumn{5}{c}{$n_t = 100$} & \multicolumn{5}{c}{$n_t = 200$} \\
\cmidrule(lr){2-6}\cmidrule(lr){7-11}
Oracle & 0.047 & 0.022 & 0.021 & 0.944 & $\diagup$ & 0.047 & 0.022 & 0.021 & 0.944 & $\diagup$ \\
DeOffline & 0.047 & 0.022 & 0.021 & 0.943 & 0.049 & 0.047 & 0.022 & 0.021 & 0.943 & 0.049 \\
LaOnline & 0.134 & 0.023 & 0.020 & 0.689 & 0.208 & 0.127 & 0.024 & 0.020 & 0.719 & 0.263 \\
DeCOER & 0.049 & 0.022 & 0.020 & 0.927 & 0.053 & 0.049 & 0.022 & 0.020 & 0.930 & 0.055 \\
\cmidrule(lr){2-6}\cmidrule(lr){7-11}
      & \multicolumn{5}{c}{$n_t = 500$} & \multicolumn{5}{c}{$n_t = 1000$} \\
\cmidrule(lr){2-6}\cmidrule(lr){7-11}
Oracle & 0.047 & 0.022 & 0.021 & 0.944 & $\diagup$ & 0.047 & 0.022 & 0.021 & 0.944 & $\diagup$ \\
DeOffline & 0.047 & 0.022 & 0.021 & 0.943 & 0.049 & 0.047 & 0.022 & 0.021 & 0.943 & 0.049 \\
LaOnline & 0.129 & 0.022 & 0.020 & 0.700 & 0.202 & 0.172 & 0.023 & 0.020 & 0.604 & 0.126 \\
DeCOER & 0.048 & 0.022 & 0.020 & 0.933 & 0.052 & 0.048 & 0.022 & 0.020 & 0.934 & 0.051 \\
\bottomrule
\end{tabular}
     \begin{tablenotes}
	\footnotesize
    \item MSE are reported times $10^{-2}$ 
    \end{tablenotes}
\label{tab:DeCOER-homo-fix}%
\vspace{12pt}
  \centering
  \renewcommand{\arraystretch}{0.6}
  \caption{Performance of Heterogeneous model, with fixed the whole sample size $N_b=100,000 (\tau=0.25)$}
\begin{tabular}{ccccccccccc}
\toprule
Method & MSE & SD & SE & CP & FPR & MSE & SD & SE & CP & FPR \\
\midrule
\multicolumn{11}{c}{$\epsilon \sim N(0,1)$} \\
\midrule
      & \multicolumn{5}{c}{$n_t = 100$} & \multicolumn{5}{c}{$n_t = 200$} \\
\cmidrule(lr){2-6}\cmidrule(lr){7-11}
Oracle & 0.122 & 0.030 & 0.030 & 0.944 & $\diagup$ & 0.122 & 0.030 & 0.030 & 0.944 & $\diagup$ \\
DeOffline & 0.121 & 0.029 & 0.029 & 0.940 & 0.050 & 0.121 & 0.029 & 0.029 & 0.940 & 0.050 \\
LaOnline & 3.836 & 0.051 & 0.029 & 0.462 & 0.172 & 3.716 & 0.046 & 0.029 & 0.470 & 0.173 \\
DeCOER & 0.135 & 0.030 & 0.029 & 0.932 & 0.051 & 0.132 & 0.029 & 0.029 & 0.932 & 0.052 \\
\cmidrule(lr){2-6}\cmidrule(lr){7-11}
      & \multicolumn{5}{c}{$n_t = 500$} & \multicolumn{5}{c}{$n_t = 1000$} \\
\cmidrule(lr){2-6}\cmidrule(lr){7-11}
Oracle & 0.122 & 0.030 & 0.030 & 0.944 & $\diagup$ & 0.122 & 0.030 & 0.030 & 0.944 & $\diagup$ \\
DeOffline & 0.121 & 0.029 & 0.029 & 0.940 & 0.050 & 0.121 & 0.029 & 0.029 & 0.940 & 0.050 \\
LaOnline & 2.670 & 0.031 & 0.029 & 0.523 & 0.195 & 1.448 & 0.034 & 0.028 & 0.619 & 0.372 \\
DeCOER & 0.131 & 0.029 & 0.029 & 0.933 & 0.052 & 0.129 & 0.029 & 0.028 & 0.930 & 0.056 \\

\midrule
\multicolumn{11}{c}{$\epsilon \sim t(3)$} \\
\midrule
      & \multicolumn{5}{c}{$n_t = 100$} & \multicolumn{5}{c}{$n_t = 200$} \\
\cmidrule(lr){2-6}\cmidrule(lr){7-11}
Oracle & 0.544 & 0.062 & 0.059 & 0.944 &  $\diagup$ & 0.544 & 0.062 & 0.059 & 0.944 & $\diagup$ \\
DeOffline & 0.538 & 0.061 & 0.058 & 0.940 & 0.051 & 0.538 & 0.061 & 0.058 & 0.940 & 0.051 \\
LaOnline & 9.936 & 0.098 & 0.056 & 0.545 & 0.286 & 9.912 & 0.102 & 0.056 & 0.538 & 0.283 \\
DeCOER & 0.580 & 0.061 & 0.056 & 0.924 & 0.056 & 0.577 & 0.061 & 0.056 & 0.927 & 0.055 \\
\cmidrule(lr){2-6}\cmidrule(lr){7-11}
      & \multicolumn{5}{c}{$n_t = 500$} & \multicolumn{5}{c}{$n_t = 1000$} \\
\cmidrule(lr){2-6}\cmidrule(lr){7-11}
Oracle & 0.544 & 0.062 & 0.059 & 0.944 & $\diagup$& 0.544 & 0.062 & 0.059 & 0.944 & $\diagup$ \\
DeOffline & 0.538 & 0.061 & 0.058 & 0.940 & 0.051 & 0.538 & 0.061 & 0.058 & 0.940 & 0.051 \\
LaOnline & 13.397 & 0.085 & 0.056 & 0.469 & 0.165 & 14.439 & 0.115 & 0.056 & 0.469 & 0.210 \\
DeCOER & 0.580 & 0.061 & 0.057 & 0.929 & 0.052 & 0.579 & 0.061 & 0.057 & 0.928 & 0.052 \\
\bottomrule
\end{tabular}
    \begin{tablenotes}
	\footnotesize
    \item MSE are reported times $10^{-2}$. 
    \end{tablenotes}
  \label{tab:DeCOER-heter-fix}
\end{table}

The above conclusions also hold under the heterogeneous model and heavy-tailed error distribution. In addition, Tables~\ref{tab:DeCOER-homo-fix-75} and~\ref{tab:DeCOER-heter-fix-75} show that when $\tau=0.75$, the performance of all methods is broadly consistent with that under $\tau=0.25$, further confirming the robustness of the proposed DeCOER method across different expectile levels and data-generating mechanisms.

\begin{table}[htbp]
  \footnotesize
  \centering
  \renewcommand{\arraystretch}{0.6}
  \caption{Performance of Homogeneous model, with fixed the batch sample size $n_t=200 (\tau=0.25)$}
\begin{tabular}{ccccccccccc}
\toprule
Method & MSE & SD & SE & CP & FPR & MSE & SD & SE & CP & FPR \\
\midrule
\multicolumn{11}{c}{$\epsilon \sim N(0,1)$} \\
\midrule
      & \multicolumn{5}{c}{$b = 10$} & \multicolumn{5}{c}{$b = 20$} \\
\cmidrule(lr){2-6}\cmidrule(lr){7-11}
Oracle & 0.056 & 0.024 & 0.024 & 0.957 & $\diagup$ & 0.027 & 0.016 & 0.017 & 0.953 & $\diagup$ \\
DeOffline & 0.058 & 0.024 & 0.024 & 0.950 & 0.051 & 0.028 & 0.017 & 0.017 & 0.950 & 0.051 \\
LaOnline & 0.185 & 0.024 & 0.023 & 0.650 & 0.139 & 0.091 & 0.017 & 0.016 & 0.658 & 0.135 \\
DeCOER & 0.058 & 0.024 & 0.023 & 0.938 & 0.057 & 0.028 & 0.017 & 0.016 & 0.945 & 0.054 \\
\cmidrule(lr){2-6}\cmidrule(lr){7-11}
      & \multicolumn{5}{c}{$b = 50$} & \multicolumn{5}{c}{$b = 100$} \\
\cmidrule(lr){2-6}\cmidrule(lr){7-11}
Oracle & 0.011 & 0.011 & 0.011 & 0.947 & $\diagup$& 0.005 & 0.007 & 0.007 & 0.954 & $\diagup$ \\
DeOffline & 0.011 & 0.011 & 0.011 & 0.948 & 0.049 & 0.005 & 0.007 & 0.007 & 0.952 & 0.050 \\
LaOnline & 0.035 & 0.011 & 0.010 & 0.678 & 0.136 & 0.018 & 0.007 & 0.007 & 0.682 & 0.138 \\
DeCOER & 0.011 & 0.011 & 0.010 & 0.939 & 0.051 & 0.006 & 0.007 & 0.007 & 0.948 & 0.050 \\

\midrule
\multicolumn{11}{c}{$\epsilon \sim t(3)$} \\
\midrule
      & \multicolumn{5}{c}{$b = 10$} & \multicolumn{5}{c}{$b = 20$} \\
\cmidrule(lr){2-6}\cmidrule(lr){7-11}
Oracle & 0.224 & 0.047 & 0.046 & 0.954 & $\diagup$& 0.105 & 0.032 & 0.033 & 0.952 & $\diagup$ \\
DeOffline & 0.224 & 0.047 & 0.045 & 0.944 & 0.052 & 0.105 & 0.032 & 0.032 & 0.947 & 0.050 \\
LaOnline & 0.938 & 0.072 & 0.043 & 0.656 & 0.236 & 0.288 & 0.035 & 0.031 & 0.708 & 0.255 \\
DeCOER & 0.242 & 0.048 & 0.042 & 0.916 & 0.066 & 0.109 & 0.032 & 0.030 & 0.935 & 0.060 \\
\cmidrule(lr){2-6}\cmidrule(lr){7-11}
      & \multicolumn{5}{c}{$b = 50$} & \multicolumn{5}{c}{$b = 100$} \\
\cmidrule(lr){2-6}\cmidrule(lr){7-11}
Oracle & 0.047 & 0.022 & 0.021 & 0.944 & $\diagup$ & 0.022 & 0.015 & 0.015 & 0.945 & $\diagup$ \\
DeOffline & 0.047 & 0.022 & 0.021 & 0.943 & 0.049 & 0.022 & 0.015 & 0.015 & 0.944 & 0.050 \\
LaOnline & 0.127 & 0.024 & 0.020 & 0.719 & 0.263 & 0.063 & 0.017 & 0.014 & 0.717 & 0.259 \\
DeCOER & 0.049 & 0.022 & 0.020 & 0.930 & 0.055 & 0.023 & 0.015 & 0.014 & 0.935 & 0.053 \\
\bottomrule
\end{tabular}
   \begin{tablenotes}
	\footnotesize
	\item  MSE are reported times $10^{-2}$.
\end{tablenotes}
  \label{tab:DeCOER-homo-stream}
\vspace{12pt}
  \centering
  \renewcommand{\arraystretch}{0.6}
  \caption{Performance of Heterogeneous model, with fixed the batch sample size $n_t=200 (\tau=0.25)$}
\begin{tabular}{ccccccccccc}
\toprule
Method & MSE & SD & SE & CP & FPR & MSE & SD & SE & CP & FPR \\
\midrule
\multicolumn{11}{c}{$\epsilon \sim N(0,1)$} \\
\midrule
      & \multicolumn{5}{c}{$b = 10$} & \multicolumn{5}{c}{$b = 20$} \\
\cmidrule(lr){2-6}\cmidrule(lr){7-11}
Oracle & 0.607 & 0.066 & 0.066 & 0.956 & $\diagup$ & 0.295 & 0.046 & 0.047 & 0.951 &$\diagup$  \\
DeOffline & 0.648 & 0.065 & 0.065 & 0.946 & 0.053 & 0.311 & 0.046 & 0.046 & 0.944 & 0.052 \\
LaOnline & 17.814 & 0.102 & 0.064 & 0.474 & 0.188 & 9.347 & 0.074 & 0.046 & 0.469 & 0.176 \\
DeCOER & 0.765 & 0.065 & 0.064 & 0.918 & 0.058 & 0.360 & 0.046 & 0.046 & 0.928 & 0.054 \\
\cmidrule(lr){2-6}\cmidrule(lr){7-11}
      & \multicolumn{5}{c}{$b = 50$} & \multicolumn{5}{c}{$b = 100$} \\
\cmidrule(lr){2-6}\cmidrule(lr){7-11}
Oracle & 0.122 & 0.030 & 0.030 & 0.944 & $\diagup$ & 0.056 & 0.020 & 0.021 & 0.954 & $\diagup$ \\
DeOffline & 0.121 & 0.029 & 0.029 & 0.940 & 0.050 & 0.056 & 0.020 & 0.021 & 0.952 & 0.049 \\
LaOnline & 3.716 & 0.046 & 0.029 & 0.470 & 0.173 & 1.834 & 0.033 & 0.021 & 0.475 & 0.181 \\
DeCOER & 0.132 & 0.029 & 0.029 & 0.932 & 0.052 & 0.060 & 0.020 & 0.021 & 0.947 & 0.050 \\

\midrule
\multicolumn{11}{c}{$\epsilon \sim t(3)$} \\
\midrule
      & \multicolumn{5}{c}{$b = 10$} & \multicolumn{5}{c}{$b = 20$} \\
\cmidrule(lr){2-6}\cmidrule(lr){7-11}
Oracle & 2.468 & 0.133 & 0.129 & 0.953 & $\diagup$ & 1.157 & 0.091 & 0.091 & 0.951 & $\diagup$ \\
DeOffline & 2.422 & 0.130 & 0.121 & 0.932 & 0.058 & 1.159 & 0.091 & 0.088 & 0.945 & 0.054 \\
LaOnline & 38.263 & 0.173 & 0.105 & 0.494 & 0.244 & 19.668 & 0.124 & 0.083 & 0.540 & 0.281 \\
DeCOER & 2.628 & 0.129 & 0.118 & 0.916 & 0.064 & 1.233 & 0.089 & 0.084 & 0.930 & 0.061 \\
\cmidrule(lr){2-6}\cmidrule(lr){7-11}
      & \multicolumn{5}{c}{$b = 50$} & \multicolumn{5}{c}{$b = 100$} \\
\cmidrule(lr){2-6}\cmidrule(lr){7-11}
Oracle & 0.544 & 0.062 & 0.059 & 0.944 & $\diagup$ & 0.217 & 0.040 & 0.041 & 0.946 & $\diagup$ \\
DeOffline & 0.538 & 0.061 & 0.058 & 0.940 & 0.051 & 0.217 & 0.040 & 0.041 & 0.946 & 0.051 \\
LaOnline & 9.912 & 0.102 & 0.056 & 0.538 & 0.283 & 5.511 & 0.071 & 0.040 & 0.520 & 0.268 \\
DeCOER & 0.577 & 0.061 & 0.056 & 0.927 & 0.055 & 0.240 & 0.040 & 0.040 & 0.933 & 0.053 \\
\bottomrule
\end{tabular}
   \begin{tablenotes}
	\footnotesize
	\item  MSE are reported times $10^{-2}$.
\end{tablenotes}
  \label{tab:DeCOER-heter-stream}
\end{table}

Tables~\ref{tab:DeCOER-homo-stream} and~\ref{tab:DeCOER-heter-stream} present the estimation performance of different methods under the setting with fixed batch size $n_t=200$. As shown, with the number of batches $b$ increases, the MSE of all methods decrease to zero, illustrate the statistical consistency of the estimators. Among them, the convergence of DeCOER is highly consistent with those of Oracle and DeOffline, indicating that the proposed DeCOER are capable for streaming data scenario. In terms of statistical inference, the SD and SE of DeCOER remain well matched. By contrast, LaOnline suffers from evident under-coverage and inflated false positive rates. Under the more challenging heavy-tailed and heteroscedastic settings, although the overall error level increases, DeCOER still maintains strong robustness, further confirming its applicability in complex streaming data environments. Tables~\ref{tab:DeCOER-homo-stream-75} and~\ref{tab:DeCOER-heter-stream-75} show that when $\tau=0.75$, the model performance is broadly similar to the above results. For brevity, we omit further discussion.

\subsection{Local Testing Power Analysis}\label{sec6-sub3}

In this section we focuses on the testing power of the DeCOER method. We consider data are generated as model defined as Eq.\eqref{eq:DeCOER-sim}, with the covariate vector is
$\boldsymbol{x}_{ti} = \{1, \sigma(\mathbf{Z}_{ti})\}
= \{1, \Phi(z_{ti,1}), \Phi(z_{ti,2}), \Phi(z_{ti,3}), z_{ti,4}, \ldots, z_{ti,p}\}$, a $(p+1)$-dimensional random vector, and $\mathbf{Z}_{ti}\sim N(0,I_p)$ are i.i.d. The random error satisfies $\epsilon_{ti}\sim N(0,1)$.

For the true parameters, we consider high-dimensional settings with $p\in\{200,400\}$ and set the number of true nonzero signals to $s=5$. The regression coefficient vector is
$\boldsymbol{\beta}^\star = \left(1, \frac{k}{\sqrt{N_b}}, 1.5, 1.5, 1.5, 1.5, 0, \ldots, 0\right)^\top$,
where the second component is used to characterize the case in which the signal strength changes with the sample size under local alternatives. To further assess the robustness of the method under different variance structures, we consider two settings:
(1) the homogeneous model, where $\boldsymbol{\gamma}=(1,0,0,\ldots,0)^\top$;
and (2) the heterogeneous model, where $\boldsymbol{\gamma}=(1,0,1,0.5,0,\ldots,0)^\top$.

Under the above model framework, we consider the following hypothesis testing problem for the parameter $\beta_1^\star(\tau)$:
\begin{equation*}\label{eq:nulltest}
H_0: \beta_1^\star(\tau) = 0
\quad \text{v.s.} \quad
H_1: \beta_1^\star(\tau) = \frac{k}{\sqrt{N_b}},
\end{equation*}
where $k\in\{0,1,\ldots,10\}$. By varying $k$, we generate a sequence of local alternatives around the null hypothesis. The test is conducted using the large-sample $z$ statistic developed in Section~\ref{sec:theory}. To evaluate Type I error control and local power, each simulation setting is repeated 200 times, and the rejection frequency is recorded. When $k=0$, this rejection frequency corresponds to the empirical Type I error rate; when $k>0$, it represents the testing power under different local signal strengths. The total sample size is set to $N_b=2000$, and the data arrive sequentially in $b=10$ streaming batches. Thus, the batch size is $n_t=N_b/b=200$.

\begin{figure}[H]
    \centering
        \includegraphics[width=.9\linewidth]{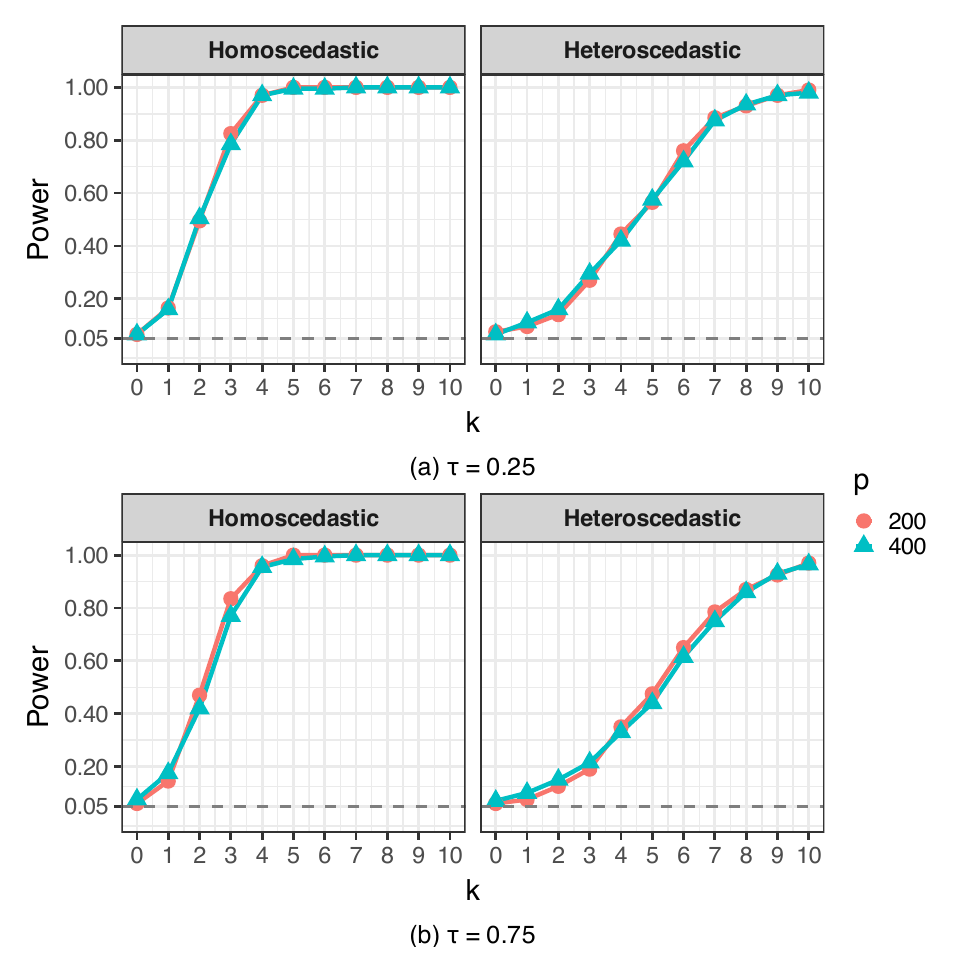}
    
    \caption{Testing power of the DeCOER method under $\tau=0.25,0.75$.}
    \label{fig:power}
\end{figure}

Figure~\ref{fig:power} displays the empirical type-I error rate and testing power of the $z$ test as the local alternative parameter $k$ varies at expectile level equal to $\tau=0.25$ and $\tau=0.75$. Overall, when $k=0$, the empirical rejection rates of the proposed method remain stable around the significance level $\alpha=0.05$, indicating that DeCOER can control the type-I error well under finite samples. As $k$ gradually increases, the testing power gradually converges to 1 under the homogeneous model. By contrast, under the heterogeneous model, the increase in testing power is relatively slower, and a larger value of $k$ is required to reach a level comparable to that under the homoscedastic setting. In addition, as the dimension $p$ increases, the overall trend of the power curves remains largely unchanged. This indicates that the proposed method remains stable in high-dimensional settings.

\section{Empirical Study}\label{sec:empirical}

\subsection{Financial and macro data}
In this section, we evaluate the proposed DeCOER method with a U.S. macroeconomic dataset, FRED-MD. This dataset contains major macroeconomic and financial indicators from January 1959 to December 2025. Detailed variable descriptions can be found in \citet{mccracken2016fred} . The goal of this study is to characterize the relationship between the S\&P 500 index and key monetary variables. Specifically in this paper we select the monetary base (BOGMBASE) and the M1 money stock (M1SL) as representative indicators. 

Since the dataset is recorded at a monthly frequency, it can naturally be regarded as streaming data arriving sequentially over time. To match the streaming data framework, this study partitions the data into local batches every 60 months, corresponding to five years, so that the batch size is $n_t=60$. To capture the dynamic effects of macroeconomic variables, third-order lagged terms of each covariate are included as explanatory variables.  After incorporating lagged variables, the covariate dimension is approximately $p\approx 360$, forming a typical high-dimensional streaming data structure.

To systematically evaluate the estimation accuracy and statistical inference performance of DeCOER, this study conducts significance tests for key monetary variables using both single-parameter Wald tests and multi-parameter joint tests. Single-variable significance tests are performed using the $z$ statistic constructed in Section~\ref{sec:theory}. For multi-parameter joint tests, we adopt a general linear restriction framework to test the joint significance of multiple lagged coefficients of the same variable. For example, for BOGMBASE, we consider the following null and alternative hypotheses:
\begin{equation*}
H_0: \mathbf{H}\tilde{\boldsymbol{\beta}} = \boldsymbol{0}
\quad \text{v.s.} \quad
H_1: \mathbf{H}\tilde{\boldsymbol{\beta}} \neq \boldsymbol{0},
\end{equation*}
where $\tilde{\boldsymbol{\beta}}$ is the debiased estimator, and
$\mathbf{H}\in\mathbb{R}^{3\times p}$ is a linear selection matrix that extracts the coefficients of the corresponding lagged terms. Denote
$\tilde{\boldsymbol{\beta}}_{\text{group}}=\mathbf{H}\tilde{\boldsymbol{\beta}}$
and
$\tilde{\boldsymbol{\Omega}}_{\text{group}}
=
\mathbf{H}\tilde{\boldsymbol{\Omega}}\mathbf{H}^\top$,
where $\tilde{\boldsymbol{\Omega}}$ is the online covariance estimator. The Wald statistic is then defined as
\begin{equation*}
W
=
n\cdot
\tilde{\boldsymbol{\beta}}_{\text{group}}^\top
\left(
\tilde{\boldsymbol{\Omega}}_{\text{group}}
\right)^{-1}
\tilde{\boldsymbol{\beta}}_{\text{group}}.
\end{equation*}
Under the null hypothesis, we have
$W \xrightarrow{d} \chi^2(3)$.

Table~\ref{tab:DeCOER-macro_full} reports the estimated coefficients and inference results for the first- to third-order lags of the two key monetary variables. The results show that the regression coefficients of BOGMBASE and M1SL vary across different expectile levels $\tau$, indicating that their effects on the S\&P 500 exhibit distributional heterogeneity. However, in most cases, the individual lag effects are not statistically significant. The joint tests further suggest that neither variable has a significant overall effect on the S\&P 500 over the full sample period.
\begin{table}[htbp]
  \centering
  \caption{Effects of Monetary Policy Variables on the S\&P 500: Full Sample (1959--2025)}
  \label{tab:DeCOER-macro_full}
  \footnotesize
  \renewcommand{\arraystretch}{0.8}
  \renewcommand\tabcolsep{4.0pt}

    \begin{tabular}{lllllll}
    \toprule
    Variable & $\tau$ & Lag 1 Coef (SE) & Lag 2 Coef (SE)  & Lag 3 Coef (SE)  & Wald Stat. & P-value \\
    \midrule
    \multirow{5}{*}{BOGMBASE} 
    & 0.1 & 0.011 (0.013) & -0.002 (0.009) & -0.010 (0.009) & 1.523 & 0.677 \\
    & 0.3 & 0.013 (0.012) & -0.001 (0.008) & -0.011 (0.008) & 2.399 & 0.494 \\
    & 0.5 & 0.016 (0.014) & -0.000 (0.007) & -0.011 (0.007)* & 3.256 & 0.354 \\
    & 0.7 & 0.019 (0.017) & 0.000 (0.008) & -0.012 (0.008)** & 3.890 & 0.274 \\
    & 0.9 & 0.022 (0.021) & 0.001 (0.009) & -0.012 (0.009)* & 3.718 & 0.294 \\
    \midrule
    \multirow{5}{*}{M1SL} 
    & 0.1 & -0.005 (0.009) & 0.004 (0.005) & -0.005 (0.005) & 2.716 & 0.438 \\
    & 0.3 & -0.015 (0.020) & 0.003 (0.004) & -0.007 (0.004)* & 3.425 & 0.331 \\
    & 0.5 & -0.016 (0.021) & 0.002 (0.002) & -0.006 (0.002) & 3.141 & 0.370 \\
    & 0.7 & -0.017 (0.022) & 0.001 (0.002) & -0.005 (0.002) & 2.856 & 0.414 \\
    & 0.9 & -0.017 (0.022) & 0.001 (0.002) & -0.005 (0.002) & 2.742 & 0.433 \\
    \bottomrule
    \end{tabular}
    \begin{tablenotes}
	   \footnotesize
        \item  Significance levels: * $p<0.1$, ** $p<0.05$, *** $p<0.01$.
    \end{tablenotes}
\end{table}

To account for the potential structural changes associated with the 2008 global financial crisis, we split the sample into the pre-crisis period (Pre-GFC, 1959--2007) and the post-crisis period (Post-GFC, 2008--2025). Tables~\ref{tab:DeCOER-macro_pre} and~\ref{tab:DeCOER-macro_post} report the results before and after the crisis, respectively. The comparison reveals clear structural differences in the effects of monetary variables on the stock market across the two periods.

In the pre-crisis period, the M1 money stock (M1SL) shows a relatively stable positive effect. Its first-order lag is statistically significant at several expectile levels, and the joint tests at the middle and lower expectiles ($\tau=0.3,0.5$) are also significant at the 10\% level. This suggests that, under the traditional monetary policy framework, liquidity indicators represented by M1 had a direct effect on the stock market.

By contrast, in the post-crisis period, the significance of M1SL weakens, while the importance of the monetary base (BOGMBASE) increases. The ER results further show that the significance of BOGMBASE is mainly concentrated in the middle and upper expectile ranges, whereas its effect is relatively weak in the lower expectiles. This may indicate that, under the prolonged accommodative monetary policy environment after the crisis, expansions in the monetary base were often associated with policy responses to potential economic risks and financial market uncertainty. Therefore, changes in the monetary base may not only reflect adjustments in liquidity supply, but also contain macroeconomic risk signals. Overall, these results suggest that the effects of monetary policy exhibit asymmetry and state dependence across different market conditions.

\begin{table}[ht]
  
  \caption{Effects of Monetary Policy Variables on the S\&P 500: Pre-GFC Period (1959--2007)}
  \label{tab:DeCOER-macro_pre}
  \footnotesize
  \centering
   \renewcommand{\arraystretch}{0.8}
    \renewcommand\tabcolsep{4.0pt}

    \begin{tabular}{lcccccc}
    \toprule
    Variable & $\tau$ & Lag 1 Coef (SE) & Lag 2 Coef (SE)  & Lag 3 Coef (SE)  & Wald Stat. & P-value \\
    \midrule
    \multirow{5}{*}{BOGMBASE} 
    & 0.1 & 0.005 (0.006) & -0.002 (0.003) & -0.001 (0.003) & 0.791 & 0.852 \\
    & 0.3 & 0.005 (0.005) & -0.002 (0.003) & -0.001 (0.003) & 0.889 & 0.828 \\
    & 0.5 & 0.004 (0.005) & -0.001 (0.003) & 0.000 (0.003) & 0.771 & 0.856 \\
    & 0.7 & 0.003 (0.005) & -0.001 (0.004) & 0.001 (0.004) & 0.487 & 0.922 \\
    & 0.9 & 0.002 (0.006) & -0.001 (0.006) & 0.001 (0.006) & 0.262 & 0.967 \\
    \midrule
    \multirow{5}{*}{M1SL} 
    & 0.1 & 0.005 (0.003)** & -0.001 (0.002) & -0.002 (0.002) & 5.478 & 0.140 \\
    & 0.3 & 0.007 (0.003)** & -0.002 (0.002) & -0.001 (0.002) & 7.204 & 0.066* \\
    & 0.5 & 0.009 (0.004)** & -0.002 (0.002) & -0.001 (0.002) & 6.881 & 0.076* \\
    & 0.7 & 0.010 (0.005)** & -0.003 (0.002) & -0.000 (0.002) & 5.786 & 0.122 \\
    & 0.9 & 0.012 (0.006)** & -0.003 (0.003) & 0.000 (0.003) & 4.772 & 0.189 \\
    \bottomrule
    \end{tabular}
         \begin{tablenotes}
	   \footnotesize
        \item  Significance levels: * $p<0.1$, ** $p<0.05$, *** $p<0.01$.
    \end{tablenotes}
\vspace{12pt}
  
  \caption{Effects of Monetary Policy Variables on the S\&P 500: Post-GFC Period (2008--2025)}
  \label{tab:DeCOER-macro_post}
   \centering
   \renewcommand{\arraystretch}{0.8}
   \renewcommand\tabcolsep{4.0pt}
    \begin{tabular}{lcccccc}
    \toprule
    Variable & $\tau$ & Lag 1 Coef (SE) & Lag 2 Coef (SE)  & Lag 3 Coef (SE)  & Wald Stat. & P-value \\
    \midrule
    \multirow{5}{*}{BOGMBASE} 
    & 0.1 & 0.015 (0.024) & -0.002 (0.017) & -0.018 (0.017) & 1.195 & 0.754 \\
    & 0.3 & 0.021 (0.023) & -0.001 (0.014) & -0.020 (0.014) & 2.156 & 0.541 \\
    & 0.5 & 0.027 (0.025) & 0.000 (0.013) & -0.021 (0.013)* & 3.384 & 0.336 \\
    & 0.7 & 0.033 (0.031) & 0.002 (0.014) & -0.023 (0.014)** & 4.715 & 0.194 \\
    & 0.9 & 0.038 (0.039) & 0.003 (0.016) & -0.024 (0.016)** & 5.327 & 0.149 \\
    \midrule
    \multirow{5}{*}{M1SL} 
    & 0.1 & -0.015 (0.017) & 0.009 (0.009) & -0.008 (0.009) & 2.381 & 0.497 \\
    & 0.3 & -0.019 (0.022) & 0.004 (0.004) & -0.007 (0.004) & 3.327 & 0.344 \\
    & 0.5 & -0.019 (0.022) & 0.002 (0.002) & -0.006 (0.002) & 3.293 & 0.349 \\
    & 0.7 & -0.019 (0.022) & 0.002 (0.002) & -0.006 (0.002) & 3.025 & 0.388 \\
    & 0.9 & -0.019 (0.023) & 0.001 (0.002) & -0.005 (0.002) & 2.828 & 0.419 \\
    \bottomrule
    \end{tabular}
         \begin{tablenotes}
	   \footnotesize
        \item  Significance levels: * $p<0.1$, ** $p<0.05$, *** $p<0.01$.
    \end{tablenotes}
\end{table}

\subsection{Gene Data}
The second dataset is the Yeast cell cycle dataset, which is available from the R package \texttt{spls}. The dataset contains gene expression information for 542 genes that are tightly regulated by the cell cycle. The goal is to predict the mRNA expression level of the target gene Alpha factor at 18 consecutive time points based on 106 chromatin immunoprecipitation (ChIP) variables corresponding to transcription factors. Focusing on both parameter inference and prediction, this study designs the following two empirical analyses:
\begin{enumerate}[label=(\arabic*)]
\item \textbf{Parameter inference}: Based on all 542 gene samples, the data are divided into five batches and perform online debiased estimation. We analysis effect of transcription factors at 18 consecutive time points and different expectile levels to characterize the effect pattern of the temporal evolution paths of key regulatory factors.

\item \textbf{Out-of-sample prediction}: We conduct 10-fold cross-validation to evaluate predictive performance. In each round, approximately 10\% of the samples are randomly assigned to the test set, while the remaining 90\% are used as the training set and further divided into five sequentially arriving streaming batches. Under this setting, we report the mean prediction expectile error (MPE) on the test set, defined as
$
\mathrm{MPE}
=
\frac{1}{n}
\sum_{i=1}^n
\rho_{\tau}
\left(
y_i-\hat{y}_i(\tau)
\right),
$
where $\hat{y}_i(\tau)$ denotes the prediction at expectile level $\tau$.
\end{enumerate}
Under the two experimental settings above, the average sample size of each data batch is approximately 100, satisfying the high-dimensional condition $n_t<p=106$. 

Table~\ref{tab:DeCOER-alpha0} presents the estimated coefficients of significant transcription factors at different expectile levels during the initial stage of the cell cycle. The results show that the sets of significant transcription factors differ substantially across expectile levels, reflecting heterogeneity in gene regulatory mechanisms. In addition, for transcription factors selected at multiple expectile levels, such as \textit{STE12} and \textit{HIR2}, their regression coefficients also vary across expectile levels, further demonstrating the necessity of distributional regression modeling.

\begin{table}[htbp]
  \caption{Estimated Coefficients of Significant Transcription Factors at the Initial Stage of the Cell Cycle}
  \label{tab:DeCOER-alpha0}
  \footnotesize
   \centering
  \renewcommand{\arraystretch}{0.6}
  \renewcommand\tabcolsep{4.0pt}
    \begin{tabular}{lrlrlrlrlr}
    \toprule
    \multicolumn{2}{c}{$\tau = 0.1$} & \multicolumn{2}{c}{$\tau = 0.3$} & \multicolumn{2}{c}{$\tau = 0.5$} & \multicolumn{2}{c}{$\tau = 0.7$} & \multicolumn{2}{c}{$\tau = 0.9$} \\
    \cmidrule(lr){1-2} \cmidrule(lr){3-4} \cmidrule(lr){5-6} \cmidrule(lr){7-8} \cmidrule(lr){9-10}
    \textit{HIR2} & -0.264 & \textit{STE12} & 0.391 & \textit{STE12} & 0.474 & \textit{STE12} & 0.554 & \textit{STE12} & 0.675 \\
    \textit{CAD1} & 0.224 & \textit{HIR2} & -0.195 & \textit{HIR2} & -0.167 & \textit{STB1} & -0.159 & \textit{MET31} & -0.203 \\
    \textit{CIN5} & 0.159 & \textit{ARG80} & -0.125 & \textit{SWI6} & -0.260 & \textit{ACE2} & 0.125 & \textit{DIG1} & -0.195 \\
    \textit{PHO4} & -0.089 & \textit{CAD1} & 0.134 & \textit{ACE2} & 0.130 & \textit{SWI6} & -0.230 & \textit{PUT3} & -0.145 \\
    \textit{REB1} & -0.133 & \textit{REB1} & -0.098 & \textit{NDD1} & -0.131 & \textit{SWI4} & -0.138 & \textit{STB1} & -0.186 \\
    \textit{ARG80} & -0.161 & \textit{SWI6} & -0.254 & \textit{STB1} & -0.119 & \textit{NDD1} & -0.164 & \textit{SWI6} & -0.275 \\
    \textit{FKH2} & -0.177 & \textit{NDD1} & -0.130 & \textit{HAA1} & 0.082 & \textit{HIR2} & -0.174 & \textit{CUP9} & 0.139 \\
    \textit{HAP4} & -0.162 & \textit{STB1} & -0.119 & \textit{ARG80} & -0.110 & \textit{IME4} & 0.120 & \textit{MET4} & 0.152 \\
    \textit{YAP7} & 0.217 & \textit{CIN5} & 0.101 & \textit{SWI4} & -0.100 & \textit{GCN4} & -0.079 & \textit{HIR2} & -0.214 \\
    \textit{STE12} & 0.262 & \textit{SWI4} & -0.115 & \textit{IME4} & 0.096 & \textit{PHO4} & -0.076 & \textit{GCN4} & -0.087 \\
    \textit{FHL1} & 0.090 & \textit{ZMS1} & -0.065 &  &  & \textit{PUT3} & -0.078 & \textit{DAL82} & -0.101 \\
    \textit{SWI6} & -0.342 & \textit{RME1} & 0.108 &  &  & \textit{HAA1} & 0.091 & \textit{SIP4} & -0.132 \\
    \textit{FKH1} & 0.088 &  &  &  &  & \textit{MET31} & -0.075 & \textit{YJL206C} & -0.142 \\
     &  &  &  &  &  & \textit{AZF1} & 0.106 & \textit{PHO4} & -0.090 \\
     &  &  &  &  &  & \textit{SIP4} & -0.094 & \textit{ACE2} & 0.103 \\
     &  &  &  &  &  & \textit{MAC1} & -0.079 & \textit{HAA1} & 0.106 \\
     &  &  &  &  &  &  &  & \textit{NDD1} & -0.159 \\
     &  &  &  &  &  &  &  & \textit{YAP3} & 0.097 \\
     &  &  &  &  &  &  &  & \textit{SWI4} & -0.115 \\
     &  &  &  &  &  &  &  & \textit{RPH1} & -0.082 \\
     &  &  &  &  &  &  &  & \textit{GAT3} & -0.158 \\
     &  &  &  &  &  &  &  & \textit{FKH2} & -0.148 \\
     &  &  &  &  &  &  &  & \textit{RME1} & 0.101 \\
     &  &  &  &  &  &  &  & \textit{IME4} & 0.130 \\
     &  &  &  &  &  &  &  & \textit{SFP1} & 0.102 \\
     &  &  &  &  &  &  &  & \textit{SFL1} & 0.077 \\
    \bottomrule
    \end{tabular}
\end{table}

\begin{figure}[htbp]
    \centering
    \includegraphics[width=.95\linewidth]{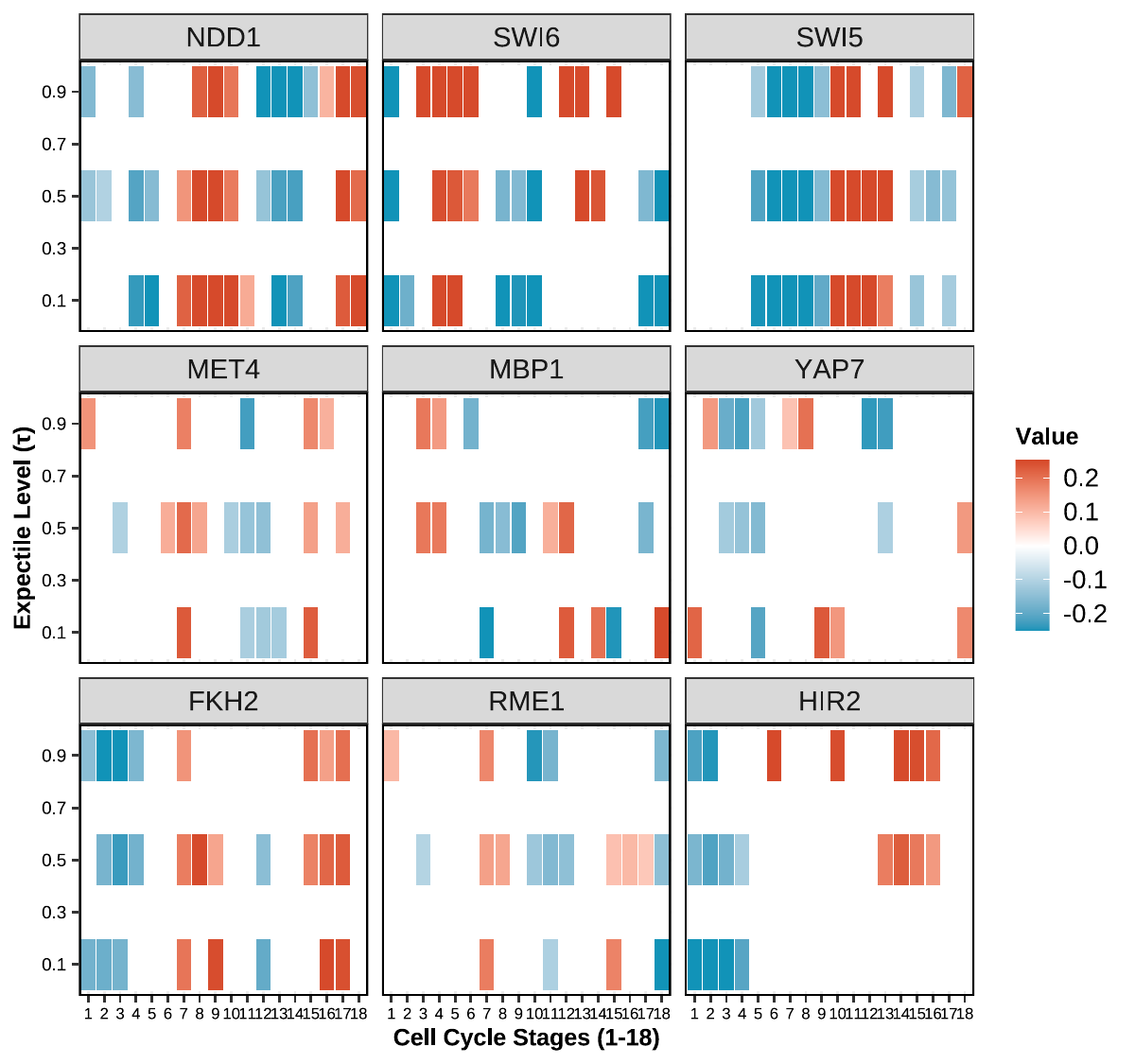}
    \caption{Top Nine Significant Transcription Factors Selected Across Different Cell Cycle Stages.}
    \label{fig:DeCOER-heat}
\end{figure}

Figure~\ref{fig:DeCOER-heat} displays the top nine selected transcription factors which are significant across different cell cycle stages (1--18) and expectile levels $\tau$. It shows that the significance of the same transcription factor exhibits dynamic patterns as the cell cycle progresses. Meanwhile, at the same cell cycle stage, the significant transcription factors also differ across $\tau$. This suggests that the regulatory effects of transcription factors on gene expression are not only time-dependent, but also vary with the location of the conditional response distribution.

\begin{figure}[htbp]
    \centering
    \includegraphics[width=1\linewidth]{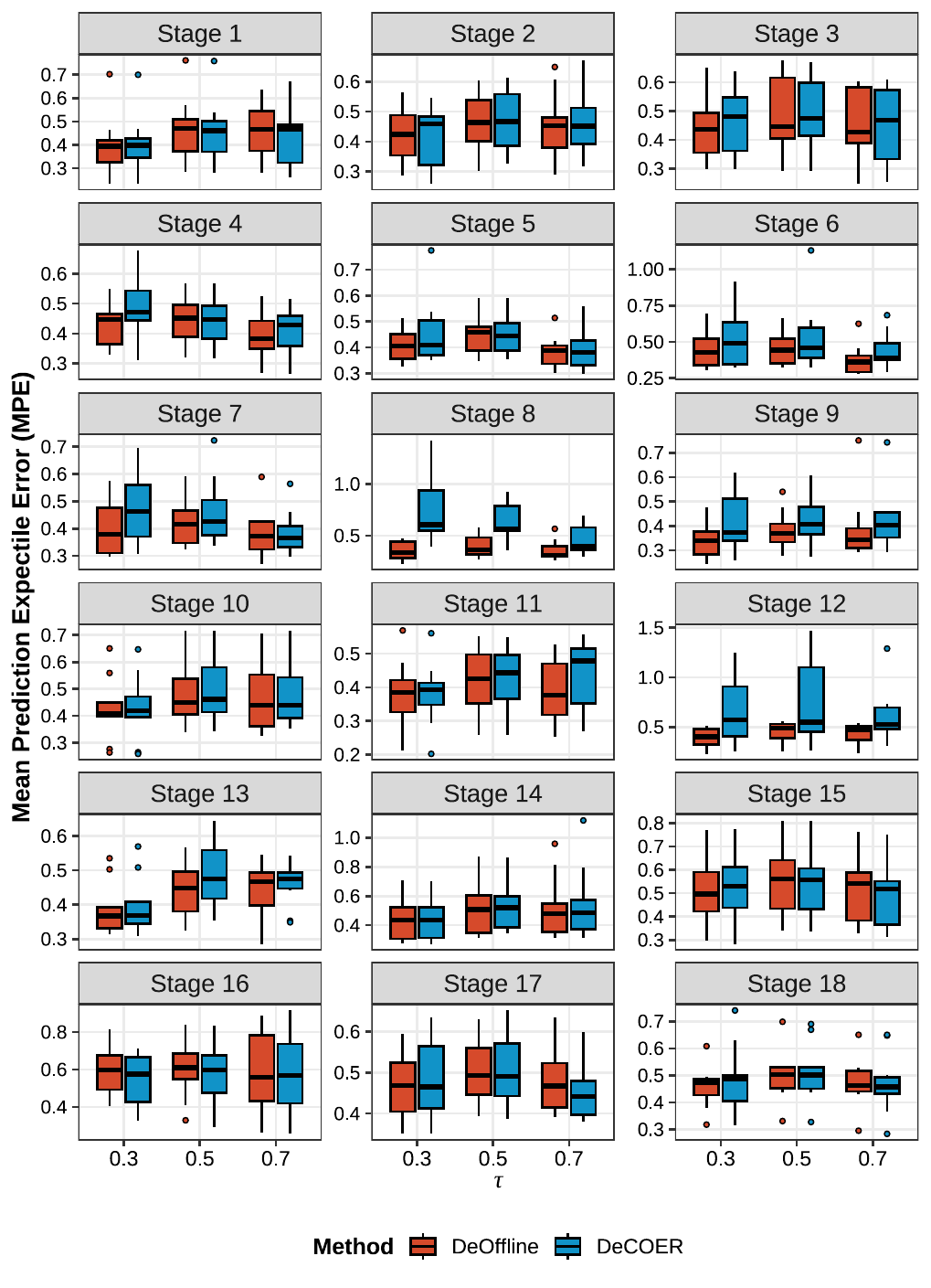}
    \caption{MPE across different cell cycle stages at expectile levels $\tau=0.3,0.5,0.7$}
    \label{fig:DeCOER-MPE}
\end{figure}

Finally, Figure~\ref{fig:DeCOER-MPE} and Table~\ref{tab:DeCOER-loss} report the average prediction loss of different methods at individual cell cycle stages and over the entire cell cycle, respectively. As shown in Figure~\ref{fig:DeCOER-MPE}, although DeCOER exhibits a larger dispersion of MPE than DeOffline at some stages (such as stage 8 or stage 12), the median MPE values of DeCOER are close to those of DeOffline in most cases, and the overall differences remain acceptable.

Furthermore, Table~\ref{tab:DeCOER-loss} shows that the overall predictive performance of DeCOER is close to that of DeOffline. This indicates that, in streaming data settings, the proposed online debiased estimation method can effectively integrate information from sequentially arriving batches and achieve prediction accuracy comparable to that of the global offline method, without storing or repeatedly accessing the complete historical data.

\begin{table}[htbp]
    \centering
    \footnotesize
  \renewcommand{\arraystretch}{0.8}
  \renewcommand\tabcolsep{4.0pt}
   
  \caption{MPE of 10-fold cv over the whole cell cycle stages}
  \label{tab:DeCOER-loss}
    \begin{tabular}{cccccccccc}
    \toprule
    \multirow{2}{*}{\textbf{Method}} & \multicolumn{9}{c}{$\tau$} \\
    \cmidrule(lr){2-10}
    & 0.1 & 0.2 & 0.3 & 0.4 & 0.5 & 0.6 & 0.7 & 0.8 & 0.9 \\
    \midrule
    DeOffline & 0.314 & 0.375 & 0.428 & 0.459 & 0.470 & 0.464 & 0.443 & 0.398 & 0.339 \\
   DeCOER & 0.500 & 0.448 & 0.486 & 0.495 & 0.511 & 0.514 & 0.467 & 0.451 & 0.483 \\
    \bottomrule
    \end{tabular}
\end{table}

\section{Conclusion and Discussion}\label{sec:conclusion}

In this paper, we addresse high-dimensional problems in streaming data and proposes a debiased renewable estimation method for high-dimensional streaming ER, referred to as DeCOER. The proposed method constructs an approximate cumulative estimating equation based on recursively updated cumulative statistics, thereby avoiding the storage and repeated access of historical data and enabling renewable parameter estimation. It further incorporates the Lasso penalty to perform variable selection in high-dimensional settings. In addition, by combining the theoretical penalty rate, this paper designs a warm-up adaptive tuning mechanism for the penalty parameter, which eliminates the computational burden of batch-wise cross-validation and effectively improves optimization efficiency. For algorithm implementation, to handle the singularity of high-dimensional covariance matrices, an online precision matrix estimation algorithm is developed. The BIC criterion is used to determine the optimal tuning parameter for nodewise regression. In addition we also systematically establishes the large-sample asymptotic properties of the DeCOER estimator and provides the corresponding online statistical inference procedure. Numerical simulations and empirical study show that the proposed method can achieve efficient parameter estimation for high-dimensional streaming data, and the debiased estimates attain accuracy comparable to that of offline full-sample estimation. 

To conclude, we acknowledge several limitations of the present work and outline directions for future research. First, the current framework assumes homogeneity across data batches. Future work may incorporate batch-level heterogeneity or change-point detection techniques to enhance the model’s robustness to abrupt structural changes. Second, due to the matrix operations involved in the debiasing procedure, the proposed method has a relatively high computational cost per update, and its efficiency may be limited when multiple expectile levels are estimated simultaneously. Future research may introduce distributed computing frameworks or expectile process estimation to enable joint estimation or mapping across multiple expectile levels, thereby improving overall computational efficiency. 
Third, under the streaming data setting, this study approximates the historical weights using the online updated estimates from the corresponding historical batches, rather than recomputing them based on the latest iterative value from the current batch. This simplification may affect the convergence rate of the estimator to some extent. Future work may consider incorporating a dynamic reconstruction strategy, in which the stored summary information from historical batches is combined with a local Taylor expansion to correct the resulting approximation error \citep{Han04062026}. Finally, the current study focuses mainly on the linear model framework. Extending the proposed method to nonlinear or semiparametric models would further broaden its applicability.

\bibliography{mybib}
\bibliographystyle{apalike}

\appendix
\setcounter{table}{0}
\setcounter{figure}{0}

\section{Technical Proofs} \label{app:proof}
\subsection{Proof of Lemma~\ref{DeCOER-L1}}\label{chap5-sec7-sub1}

Let $\boldsymbol{\Delta}_b=\ddot{\boldsymbol{\beta}}_b-\boldsymbol{\beta}_0$. Since
$\ddot{\boldsymbol{\beta}}_b$ is the ideal minimizer of the Lasso-penalized surrogate objective
$\mathcal Q_{b,\lambda}^{\mathrm{s}}$, the following basic inequality holds:
\[
\mathcal Q_b^{\mathrm{s}}(\boldsymbol{\beta}_0+\boldsymbol{\Delta}_b)
+\lambda_b\|\boldsymbol{\beta}_0+\boldsymbol{\Delta}_b\|_1
\le
\mathcal Q_b^{\mathrm{s}}(\boldsymbol{\beta}_0)
+\lambda_b\|\boldsymbol{\beta}_0\|_1 .
\]
Let $R_b(\boldsymbol{\Delta}_b)$ denote the first-order remainder of the surrogate objective expanded at $\boldsymbol{\beta}_0$ along the direction $\boldsymbol{\Delta}_b$, namely
\begin{equation*}
R_b(\boldsymbol{\Delta}_b)
=
\mathcal Q_b^{\mathrm{s}}(\boldsymbol{\beta}_0+\boldsymbol{\Delta}_b)
-
\mathcal Q_b^{\mathrm{s}}(\boldsymbol{\beta}_0)
-
\nabla\mathcal Q_b^{\mathrm{s}}(\boldsymbol{\beta}_0)^\top
\boldsymbol{\Delta}_b .
\end{equation*}
By Condition~\ref{con:chap5-8} and H\"older's inequality,
\[
-\nabla\mathcal Q_b^{\mathrm{s}}(\boldsymbol{\beta}_0)^\top
\boldsymbol{\Delta}_b
\le
\frac{\lambda_b}{2}
\|\boldsymbol{\Delta}_b\|_1 .
\]
Since $\boldsymbol{\beta}_{0,S^c}=\boldsymbol{0}$, we also have
\[
\|\boldsymbol{\beta}_0+\boldsymbol{\Delta}_b\|_1
-
\|\boldsymbol{\beta}_0\|_1
\ge
\|\boldsymbol{\Delta}_{b,S^c}\|_1
-
\|\boldsymbol{\Delta}_{b,S}\|_1 .
\]
Substituting the above inequalities into the basic inequality gives
\begin{equation}\label{eq:DeCOER-cone}
R_b(\boldsymbol{\Delta}_b)
+
\frac{\lambda_b}{2}
\|\boldsymbol{\Delta}_{b,S^c}\|_1
\le
\frac{3\lambda_b}{2}
\|\boldsymbol{\Delta}_{b,S}\|_1 .
\end{equation}
Since the smooth surrogate loss $\mathcal Q_b^{\mathrm{s}}$ is convex, we have
$R_b(\boldsymbol{\Delta}_b)\ge0$. Therefore,
\[
\|\boldsymbol{\Delta}_{b,S^c}\|_1
\le
3\|\boldsymbol{\Delta}_{b,S}\|_1,
\]
which means that the estimation error lies in the corresponding sparse cone. By Condition~\ref{con:chap5-8}, there exists a constant $\kappa>0$ such that
\[
R_b(\boldsymbol{\Delta}_b)
\ge
\kappa\|\boldsymbol{\Delta}_b\|_2^2 .
\]
Combining this with Eq.~\eqref{eq:DeCOER-cone} and the Cauchy--Schwarz inequality yields
\[
\kappa\|\boldsymbol{\Delta}_b\|_2^2
\le
\frac{3\lambda_b}{2}
\|\boldsymbol{\Delta}_{b,S}\|_1
\le
\frac{3\lambda_b\sqrt{s}}{2}
\|\boldsymbol{\Delta}_b\|_2 .
\]
Hence,
\[
\|\boldsymbol{\Delta}_b\|_2
=
O_p(\sqrt{s}\lambda_b).
\]
Furthermore, by the cone constraint,
\begin{equation*}
\begin{aligned}
\|\boldsymbol{\Delta}_b\|_1
&=
\|\boldsymbol{\Delta}_{b,S}\|_1
+
\|\boldsymbol{\Delta}_{b,S^c}\|_1 \\
&\le
4\|\boldsymbol{\Delta}_{b,S}\|_1
\le
4\sqrt{s}\|\boldsymbol{\Delta}_b\|_2
=
O_p(s\lambda_b).
\end{aligned}
\end{equation*}
Therefore, if $s\lambda_b\to0$, then
$\|\boldsymbol{\Delta}_b\|_1=o_p(1)$.

\subsection{Proof of Lemma~\ref{DeCOER-L2}}\label{chap5-sec7-sub2}

Let $\boldsymbol{\Delta}=\boldsymbol{\beta}_1-\boldsymbol{\beta}_2$. By Eq.~\eqref{eq:DeCOER-proxy-stats}, the difference between the gradients of the surrogate objective at
$\boldsymbol{\beta}_1$ and $\boldsymbol{\beta}_2$ is
\begin{equation*}
\begin{aligned}
&
\nabla\mathcal Q_b^{\mathrm{s}}(\boldsymbol{\beta}_1)
-
\nabla\mathcal Q_b^{\mathrm{s}}(\boldsymbol{\beta}_2) \\
=&
\frac{1}{N_b}\mathbf Z_{N_{b-1}}\boldsymbol{\Delta}
+
\frac{1}{N_b}
\mathbf X^{(b)\top}
\Bigl\{
\mathbf W_{\boldsymbol{\beta}_1}^{(b)}
(\mathbf X^{(b)}\boldsymbol{\beta}_1-\boldsymbol Y^{(b)})
-
\mathbf W_{\boldsymbol{\beta}_2}^{(b)}
(\mathbf X^{(b)}\boldsymbol{\beta}_2-\boldsymbol Y^{(b)})
\Bigr\}.
\end{aligned}
\end{equation*}
Adding and subtracting
$\mathbf W_{\boldsymbol{\beta}_2}^{(b)}
(\mathbf X^{(b)}\boldsymbol{\beta}_1-\boldsymbol Y^{(b)})$
on the right-hand side gives
\begin{equation*}
\begin{aligned}
\nabla\mathcal Q_b^{\mathrm{s}}(\boldsymbol{\beta}_1)
-
\nabla\mathcal Q_b^{\mathrm{s}}(\boldsymbol{\beta}_2)
=
\boldsymbol{\Sigma}_b(\boldsymbol{\beta}_2)\boldsymbol{\Delta}
+
\frac{1}{N_b}
\mathbf X^{(b)\top}
(\mathbf W_{\boldsymbol{\beta}_1}^{(b)}-\mathbf W_{\boldsymbol{\beta}_2}^{(b)})
(\mathbf X^{(b)}\boldsymbol{\beta}_1-\boldsymbol Y^{(b)}).
\end{aligned}
\end{equation*}
By the definition of the remainder term in the lemma,
\[
\boldsymbol r_{g,b}(\boldsymbol{\beta}_1,\boldsymbol{\beta}_2)
=
\frac{1}{N_b}
\mathbf X^{(b)\top}
(\mathbf W_{\boldsymbol{\beta}_1}^{(b)}-\mathbf W_{\boldsymbol{\beta}_2}^{(b)})
(\mathbf X^{(b)}\boldsymbol{\beta}_1-\boldsymbol Y^{(b)}).
\]
The weights $w_{\boldsymbol{\beta}_1,bi}$ and $w_{\boldsymbol{\beta}_2,bi}$ can differ only when the two fitted residuals have different signs. Therefore,
\begin{equation*}
|w_{\boldsymbol{\beta}_1,bi}-w_{\boldsymbol{\beta}_2,bi}|
\le
\boldsymbol{I}
\left\{
|y_{bi}-\boldsymbol x_{bi}^{\top}\boldsymbol{\beta}_2|
\le
c_3\|\boldsymbol{\Delta}\|_1
\right\}.
\end{equation*}
It follows that the conditional expectation of the product of the weight difference and the residual is of order
$O(\|\boldsymbol{\Delta}\|_1^2)$. Combining the local boundedness of the error density and the boundedness of covariates in Conditions~\ref{con:chap5-2}--\ref{con:chap5-3}, we obtain
\[
\|\boldsymbol r_{g,b}(\boldsymbol{\beta}_1,\boldsymbol{\beta}_2)\|_\infty
=
O_p(\|\boldsymbol{\Delta}\|_1^2).
\]

\subsection{Proof of Lemma~\ref{DeCOER-L3}}\label{chap5-sec7-sub3}

To prove the approximate inverse properties of the precision matrix estimators constructed from both the ideal estimator and the actual PGD estimator in a unified way, let
$(\boldsymbol{\Sigma}_b^*,\boldsymbol{\Theta}_b^*)$
denote either
$(\ddot{\boldsymbol{\Sigma}}_b,\ddot{\boldsymbol{\Theta}}_b)$
or
$(\widetilde{\boldsymbol{\Sigma}}_b,\widetilde{\boldsymbol{\Theta}}_b)$.
By Condition~\ref{con:chap5-6}, the nodewise scaling term satisfies
\[
(\hat d_{b,j}^*)^2
=
\sigma_{b,jj}^*
-
\boldsymbol\sigma_{b,j,-j}^{*\top}
\hat{\boldsymbol\phi}_{b,j}^*,
\]
and is bounded away from zero uniformly.

For the $j$-th node, the KKT condition gives
\begin{equation*}
\boldsymbol{\Sigma}_{b,-j,-j}^{*}
\hat{\boldsymbol\phi}_{b,j}^{*}
-
\boldsymbol\sigma_{b,-j,j}^{*}
+
\widehat\lambda_{b,j}^{*,\mathrm{node}}
\hat{\boldsymbol\kappa}_{b,j}^{*}
=
\boldsymbol{0},
\qquad
\|\hat{\boldsymbol\kappa}_{b,j}^{*}\|_\infty\le1.
\end{equation*}
Using the definitions of the scaling term and the $j$-th row of the precision matrix estimator, dividing the KKT condition by
$(\hat d_{b,j}^{*})^2$ and combining it with the $j$-th coordinate yield
\begin{equation*}
\left\|
\boldsymbol e_j^\top
-
\boldsymbol{\Theta}_{b,j}^{*}
\boldsymbol{\Sigma}_b^*
\right\|_\infty
\le
\widehat\lambda_{b,j}^{*,\mathrm{node}}
/
(\hat d_{b,j}^{*})^2 .
\end{equation*}
Taking the maximum over $j$ and using the positive lower bound of the scaling terms in Condition~\ref{con:chap5-6}, we obtain
\begin{equation*}
\left\|
\mathbf I
-
\boldsymbol{\Theta}_b^*
\boldsymbol{\Sigma}_b^*
\right\|_{\max}
\le
\max_j
\frac{
\widehat\lambda_{b,j}^{*,\mathrm{node}}
}{
(\hat d_{b,j}^{*})^2
}
=
O_p(\lambda_b^{\mathrm{node}}),
\end{equation*}
and
\begin{equation*}
\max_j\|\boldsymbol{\Theta}_{b,j}^{*}\|_1
=
\max_j
(\hat d_{b,j}^{*})^{-2}
(1+\|\hat{\boldsymbol\phi}_{b,j}^{*}\|_1)
=
O_p(1).
\end{equation*}
This proves the approximate inverse property and the row-wise $\ell_1$ boundedness of
$\boldsymbol{\Theta}_b^*$.

Next, we prove the consistency of the $j$-th row of the precision matrix estimator. The nodewise regression bias satisfies
\begin{equation*}
\begin{aligned}
\boldsymbol{\Sigma}_{b,-j,-j}^{*}\boldsymbol\phi_{0,j}
-
\boldsymbol\sigma_{b,-j,j}^{*}
=
(\boldsymbol{\Sigma}_{b,-j,-j}^{*}-\boldsymbol{\Sigma}_{-j,-j})
\boldsymbol\phi_{0,j}
-
(\boldsymbol\sigma_{b,-j,j}^{*}-\boldsymbol\sigma_{-j,j}).
\end{aligned}
\end{equation*}
By Conditions~\ref{con:chap5-7}--\ref{con:chap5-8},
\begin{equation*}
\begin{aligned}
\left\|
\boldsymbol{\Sigma}_{b,-j,-j}^{*}\boldsymbol\phi_{0,j}
-
\boldsymbol\sigma_{b,-j,j}^{*}
\right\|_\infty
&\le
\|\boldsymbol{\Sigma}_b^*-\boldsymbol{\Sigma}\|_{\max}
(1+\|\boldsymbol\phi_{0,j}\|_1) \\
&=
O_p
\left(
\sqrt{\frac{\log p}{N_b}}
+
s\lambda_b
\right)
=
O_p(\lambda_b^{\mathrm{node}}).
\end{aligned}
\end{equation*}
Together with the nodewise sparsity and restricted eigenvalue condition in Condition~\ref{con:chap5-7}, this yields
\[
\|\hat{\boldsymbol\phi}_{b,j}^{*}-\boldsymbol\phi_{0,j}\|_1
=
O_p(s_{\Theta,j}\lambda_b^{\mathrm{node}}).
\]

For the scaling term, the error between the estimator and its population counterpart can be decomposed as
\begin{equation*}
\begin{aligned}
(\hat d_{b,j}^*)^2-d_{0,j}^2
={}&
(\sigma_{b,jj}^*-\sigma_{jj})
-
(\boldsymbol\sigma_{b,j,-j}^*-\boldsymbol\sigma_{j,-j})^\top
\boldsymbol\phi_{0,j}
-
\boldsymbol\sigma_{b,j,-j}^{*\top}
(\hat{\boldsymbol\phi}_{b,j}^*-\boldsymbol\phi_{0,j}).
\end{aligned}
\end{equation*}
Using
$\|\boldsymbol{\Sigma}_b^*-\boldsymbol{\Sigma}\|_{\max}$,
$\|\boldsymbol\phi_{0,j}\|_1=O(1)$,
and the nodewise coefficient error bound, we obtain
\[
\left|
(\hat d_{b,j}^*)^2-d_{0,j}^2
\right|
=
O_p(s_{\Theta,j}\lambda_b^{\mathrm{node}}).
\]
Therefore, there exists a constant $c_5>0$ such that
\begin{equation*}
\begin{aligned}
\|
\widetilde{\boldsymbol{\Theta}}_{\tilde{\boldsymbol{\beta}}_b,j}
-
\boldsymbol{\Theta}_j
\|_1
&\le
c_5
\|\hat{\boldsymbol{\phi}}_{b,j}-\boldsymbol{\phi}_{0,j}\|_1
+
c_5
|\hat d_{b,j}^{-2}-d_{0,j}^{-2}| \\
&=
O_p(s_{\Theta,j}\lambda_b^\mathrm{node})
=
o_p(1).
\end{aligned}
\end{equation*}

\subsection{Proof of Theorem~\ref{DeCOER-T1}}\label{chap5-sec7-sub4}

Let
$\boldsymbol{\Delta}_b=\ddot{\boldsymbol{\beta}}_b-\boldsymbol{\beta}_0$,
and define the nonlinear remainder of the gradient difference as
\begin{equation*}
\boldsymbol\zeta_b
:=
\nabla\mathcal Q_b^{\mathrm{s}}(\ddot{\boldsymbol{\beta}}_b)
-
\nabla\mathcal Q_b^{\mathrm{s}}(\boldsymbol{\beta}_0)
-
\ddot{\boldsymbol{\Sigma}}_b\boldsymbol{\Delta}_b.
\end{equation*}
By Lemma~\ref{DeCOER-L2}, taking
$\boldsymbol{\beta}_1=\boldsymbol{\beta}_0$ and
$\boldsymbol{\beta}_2=\ddot{\boldsymbol{\beta}}_b$ gives
$\boldsymbol\zeta_b
=
-\boldsymbol r_{g,b}
(\boldsymbol{\beta}_0,\ddot{\boldsymbol{\beta}}_b)$.
Hence,
\[
\|\boldsymbol\zeta_b\|_\infty
=
O_p(\|\boldsymbol{\Delta}_b\|_1^2).
\]

On the other hand, by Eq.~\eqref{eq:DeCOER-proxy-stats},
\begin{equation*}
\begin{aligned}
-\nabla\mathcal Q_b^{\mathrm{s}}(\boldsymbol{\beta}_0)
=
\frac1{N_b}
\sum_{t=1}^b
\mathbf X^{(t)\top}
\mathbf W_{\boldsymbol{\beta}_0}^{(t)}
\boldsymbol\epsilon^{(t)}
+
\frac1{N_b}
\sum_{t=1}^{b-1}
\mathbf X^{(t)\top}
\left(
\mathbf W_{\tilde{\boldsymbol{\beta}}_t}^{(t)}
-
\mathbf W_{\boldsymbol{\beta}_0}^{(t)}
\right)
\boldsymbol\epsilon^{(t)}.
\end{aligned}
\end{equation*}
Substituting the above display into Eq.~\eqref{eq:DeCOER-debias-ideal}, we obtain
\[
\ddot{\boldsymbol{\beta}}_{\mathrm{de}}
-
\boldsymbol{\beta}_0
=
\ddot{\boldsymbol{\Theta}}_b
\frac1{N_b}
\sum_{t=1}^b
\mathbf X^{(t)\top}
\mathbf W_{\boldsymbol{\beta}_0}^{(t)}
\boldsymbol\epsilon^{(t)}
+
\boldsymbol r_b,
\]
where
\begin{equation*}
\begin{aligned}
\boldsymbol r_b
=
&\ddot{\boldsymbol{\Theta}}_b
\frac1{N_b}
\sum_{t=1}^{b-1}
\mathbf X^{(t)\top}
(\mathbf W_{\tilde{\boldsymbol{\beta}}_t}^{(t)}
-
\mathbf W_{\boldsymbol{\beta}_0}^{(t)})
\boldsymbol\epsilon^{(t)} \\
&+
(\mathbf I-\ddot{\boldsymbol{\Theta}}_b\ddot{\boldsymbol{\Sigma}}_b)
\boldsymbol{\Delta}_b
-
\ddot{\boldsymbol{\Theta}}_b\boldsymbol\zeta_b .
\end{aligned}
\end{equation*}
By Condition~\ref{con:chap5-13} and Lemma~\ref{DeCOER-L3}, the infinity norm of the first term is
$o_p(N_b^{-1/2})$. For the second term, by Condition~\ref{con:chap5-9},
\[
\|
(\mathbf I-\ddot{\boldsymbol{\Theta}}_b\ddot{\boldsymbol{\Sigma}}_b)
\boldsymbol{\Delta}_b
\|_\infty
\le
\|
\mathbf I-\ddot{\boldsymbol{\Theta}}_b\ddot{\boldsymbol{\Sigma}}_b
\|_{\max}
\|\boldsymbol{\Delta}_b\|_1
=
O_p(\lambda_b^{\mathrm{node}}s\lambda_b)
=
o_p(N_b^{-1/2}).
\]
Finally, by Lemmas~\ref{DeCOER-L1}--\ref{DeCOER-L3} and Condition~\ref{con:chap5-9},
\[
\|\ddot{\boldsymbol{\Theta}}_b\boldsymbol\zeta_b\|_\infty
=
O_p(\|\boldsymbol{\Delta}_b\|_1^2)
=
O_p(s^2\lambda_b^2)
=
o_p(N_b^{-1/2}).
\]
Therefore,
\[
\|\boldsymbol r_b\|_\infty
=
o_p(N_b^{-1/2}).
\]

\subsection{Proof of Theorem~\ref{DeCOER-T2}}\label{chap5-sec7-sub5}

By Theorem~\ref{DeCOER-T1}, for any fixed coordinate $j$,
\begin{equation*}
\sqrt{N_b}
(\ddot{\beta}_{\mathrm{de},j}-\beta_{0j})
=
\frac1{\sqrt{N_b}}
\sum_{t=1}^b
\sum_{i=1}^{n_t}
\ddot{\boldsymbol{\Theta}}_{b,j}
w_{\boldsymbol{\beta}_0,ti}
\epsilon_{ti}
\boldsymbol x_{ti}
+
o_p(1).
\end{equation*}
Furthermore, by Lemma~\ref{DeCOER-L3}, Condition~\ref{con:chap5-1}, and Condition~\ref{con:chap5-7},
\begin{equation*}
\frac1{\sqrt{N_b}}
\sum_{t=1}^b
\sum_{i=1}^{n_t}
(\ddot{\boldsymbol{\Theta}}_{b,j}-\boldsymbol{\Theta}_j)
w_{\boldsymbol{\beta}_0,ti}
\epsilon_{ti}
\boldsymbol x_{ti}
=
o_p(1).
\end{equation*}
Hence,
\[
\sqrt{N_b}
(\ddot{\beta}_{\mathrm{de},j}-\beta_{0j})
=
\frac1{\sqrt{N_b}}
\sum_{t=1}^b
\sum_{i=1}^{n_t}
\boldsymbol{\Theta}_j
w_{\boldsymbol{\beta}_0,ti}
\epsilon_{ti}
\boldsymbol x_{ti}
+
o_p(1).
\]
By the Lindeberg condition in Condition~\ref{con:chap5-10} and Slutsky's theorem, we have
\[
\sqrt{N_b}
(\ddot{\beta}_{\mathrm{de},j}-\beta_{0j})
\xrightarrow{d}
N(0,\Omega_{jj}).
\]

\subsection{Proof of Theorem~\ref{DeCOER-T3}}\label{chap5-sec7-sub6}

Let
$\boldsymbol e_b=\tilde{\boldsymbol{\beta}}_b-\ddot{\boldsymbol{\beta}}_b$
and
$d_{b,0}
=
\|\tilde{\boldsymbol{\beta}}_{b-1}-\ddot{\boldsymbol{\beta}}_b\|_1$.
By Condition~\ref{con:chap5-11},
\[
\|\boldsymbol e_b\|_1
\le
q^{K_b}
\|\tilde{\boldsymbol{\beta}}_{b-1}-\ddot{\boldsymbol{\beta}}_b\|_1
=
\delta_b .
\]
Combining Condition~\ref{con:chap5-12} with Lemma~\ref{DeCOER-L1}, we obtain
\begin{equation*}
\begin{aligned}
\|\tilde{\boldsymbol{\beta}}_b-\boldsymbol{\beta}_0\|_1
&\le
\|\tilde{\boldsymbol{\beta}}_b-\ddot{\boldsymbol{\beta}}_b\|_1
+
\|\ddot{\boldsymbol{\beta}}_b-\boldsymbol{\beta}_0\|_1 \\
&=
o_p(s\lambda_b)
+
O_p(s\lambda_b)
=
O_p(s\lambda_b).
\end{aligned}
\end{equation*}
Define
$\boldsymbol\xi_b
:=
-\boldsymbol r_{g,b}
(\ddot{\boldsymbol{\beta}}_b,\tilde{\boldsymbol{\beta}}_b)$.
By Lemma~\ref{DeCOER-L2},
\[
\|\boldsymbol\xi_b\|_\infty
=
O_p(\|\boldsymbol e_b\|_1^2).
\]

Since $\ddot{\boldsymbol{\beta}}_b$ is the ideal minimizer of the recursive surrogate objective
$\mathcal Q_{b,\lambda}^{\mathrm{s}}$ with the $\ell_1$ penalty, its KKT condition gives
\[
\|\nabla\mathcal Q_b^{\mathrm{s}}(\ddot{\boldsymbol{\beta}}_b)\|_\infty
\le
\lambda_b.
\]
Combining Eq.~\eqref{eq:debias-stream} with Eq.~\eqref{eq:DeCOER-debias-ideal}, we obtain
\begin{equation*}
\begin{aligned}
\tilde{\boldsymbol{\beta}}_{\mathrm{de}}
-
\ddot{\boldsymbol{\beta}}_{\mathrm{de}}
=
&
(\mathbf I-\widetilde{\boldsymbol{\Theta}}_b
\widetilde{\boldsymbol{\Sigma}}_b)
\boldsymbol e_b
-
\widetilde{\boldsymbol{\Theta}}_b
\boldsymbol\xi_b \\
&-
(\widetilde{\boldsymbol{\Theta}}_b-\ddot{\boldsymbol{\Theta}}_b)
\nabla\mathcal Q_b^{\mathrm{s}}(\ddot{\boldsymbol{\beta}}_b).
\end{aligned}
\end{equation*}
By Lemma~\ref{DeCOER-L3} and Condition~\ref{con:chap5-12},
\[
\sqrt{N_b}
\|
\tilde{\boldsymbol{\beta}}_{\mathrm{de}}
-
\ddot{\boldsymbol{\beta}}_{\mathrm{de}}
\|_\infty
=
o_p(1).
\]
Combining Condition~\ref{con:chap5-13}, Theorem~\ref{DeCOER-T2}, and Slutsky's theorem gives
\[
\sqrt{N_b}
\left(
\tilde\beta_{\mathrm{de},j}
-
\beta_{0j}
\right)
\xrightarrow{d}
N(0,\Omega_{jj}).
\]
\newpage
\section{Additional Simulation Results} \label{app:add-table}

\begin{table}[htbp]
\vspace{-0.2cm}
\footnotesize
  \centering
  \renewcommand{\arraystretch}{0.6}
  \caption {
Performance of Homogeneous model, with fixed the whole sample size $N_b=100,000 (\tau=0.75)$}
\begin{tabular}{ccccccccccc}
\toprule
Method & MSE & SD & SE & CP & FPR & MSE & SD & SE & CP & FPR \\
\midrule
\multicolumn{11}{c}{$\epsilon \sim \mathcal{N}(0,1)$} \\
\midrule
      & \multicolumn{5}{c}{$n_t = 100$} & \multicolumn{5}{c}{$n_t = 200$} \\
\cmidrule(lr){2-6}\cmidrule(lr){7-11}
Oracle & 0.011 & 0.011 & 0.011 & 0.947 & $\diagup$ & 0.011 & 0.011 & 0.011 & 0.947 & $\diagup$ \\
DeOffline & 0.011 & 0.011 & 0.011 & 0.948 & 0.048 & 0.011 & 0.011 & 0.011 & 0.948 & 0.048 \\
LaOnline & 0.067 & 0.011 & 0.011 & 0.419 & 0.024 & 0.035 & 0.011 & 0.010 & 0.679 & 0.135 \\
DeCORER & 0.012 & 0.011 & 0.011 & 0.948 & 0.046 & 0.012 & 0.011 & 0.010 & 0.943 & 0.050 \\
\cmidrule(lr){2-6}\cmidrule(lr){7-11}
      & \multicolumn{5}{c}{$n_t = 500$} & \multicolumn{5}{c}{$n_t = 1000$} \\
\cmidrule(lr){2-6}\cmidrule(lr){7-11}
Oracle & 0.011 & 0.011 & 0.011 & 0.947 & $\diagup$ & 0.011 & 0.011 & 0.011 & 0.947 & $\diagup$ \\
DeOffline & 0.011 & 0.011 & 0.011 & 0.948 & 0.048 & 0.011 & 0.011 & 0.011 & 0.948 & 0.048 \\
LaOnline & 0.031 & 0.011 & 0.010 & 0.744 & 0.234 & 0.038 & 0.011 & 0.010 & 0.648 & 0.122 \\
DeCORER & 0.012 & 0.011 & 0.010 & 0.942 & 0.052 & 0.012 & 0.011 & 0.010 & 0.939 & 0.050 \\

\midrule
\multicolumn{11}{c}{$\epsilon \sim t(3)$} \\
\midrule
      & \multicolumn{5}{c}{$n_t = 100$} & \multicolumn{5}{c}{$n_t = 200$} \\
\cmidrule(lr){2-6}\cmidrule(lr){7-11}
Oracle & 0.045 & 0.021 & 0.021 & 0.948 & $\diagup$ & 0.045 & 0.021 & 0.021 & 0.948 & $\diagup$ \\
DeOffline & 0.045 & 0.021 & 0.021 & 0.949 & 0.050 & 0.045 & 0.021 & 0.021 & 0.949 & 0.050 \\
LaOnline & 0.141 & 0.023 & 0.020 & 0.688 & 0.204 & 0.125 & 0.023 & 0.020 & 0.700 & 0.257 \\
DeCORER & 0.047 & 0.021 & 0.020 & 0.936 & 0.053 & 0.047 & 0.021 & 0.020 & 0.933 & 0.055 \\
\cmidrule(lr){2-6}\cmidrule(lr){7-11}
      & \multicolumn{5}{c}{$n_t = 500$} & \multicolumn{5}{c}{$n_t = 1000$} \\
\cmidrule(lr){2-6}\cmidrule(lr){7-11}
Oracle & 0.045 & 0.021 & 0.021 & 0.948 & $\diagup$ & 0.045 & 0.021 & 0.021 & 0.948 & $\diagup$ \\
DeOffline & 0.045 & 0.021 & 0.021 & 0.949 & 0.050 & 0.045 & 0.021 & 0.021 & 0.949 & 0.050 \\
LaOnline & 0.126 & 0.022 & 0.020 & 0.705 & 0.206 & 0.164 & 0.023 & 0.020 & 0.610 & 0.145 \\
DeCORER & 0.046 & 0.021 & 0.020 & 0.938 & 0.053 & 0.046 & 0.021 & 0.020 & 0.938 & 0.053 \\
\bottomrule
\end{tabular}
  \begin{tablenotes}
	\footnotesize
	\item MSE are reported times $10^{-2}$.
\end{tablenotes}
  \label{tab:DeCOER-homo-fix-75}
\vspace{12pt}
  \centering
   \renewcommand{\arraystretch}{0.6}
  \caption{
Performance of Heterogeneous model, with fixed the whole sample size $N_b=100,000 (\tau=0.75)$}
  
\begin{tabular}{ccccccccccc}
\toprule
Method & MSE & SD & SE & CP & FPR & MSE & SD & SE & CP & FPR \\
\midrule
\multicolumn{11}{c}{$\epsilon \sim \mathcal{N}(0,1)$} \\
\midrule
      & \multicolumn{5}{c}{$n_t = 100$} & \multicolumn{5}{c}{$n_t = 200$} \\
\cmidrule(lr){2-6}\cmidrule(lr){7-11}
Oracle & 0.126 & 0.030 & 0.030 & 0.949 & $\diagup$ & 0.126 & 0.030 & 0.030 & 0.949 & $\diagup$ \\
DeOffline & 0.129 & 0.030 & 0.030 & 0.950 & 0.050 & 0.129 & 0.030 & 0.030 & 0.950 & 0.050 \\
LaOnline & 7.126 & 0.067 & 0.031 & 0.348 & 0.101 & 3.229 & 0.050 & 0.030 & 0.502 & 0.218 \\
DeCORER & 0.134 & 0.031 & 0.031 & 0.952 & 0.045 & 0.130 & 0.030 & 0.030 & 0.947 & 0.050 \\
\cmidrule(lr){2-6}\cmidrule(lr){7-11}
      & \multicolumn{5}{c}{$n_t = 500$} & \multicolumn{5}{c}{$n_t = 1000$} \\
\cmidrule(lr){2-6}\cmidrule(lr){7-11}
Oracle & 0.126 & 0.030 & 0.030 & 0.949 & $\diagup$ & 0.126 & 0.030 & 0.030 & 0.949 & $\diagup$ \\
DeOffline & 0.129 & 0.030 & 0.030 & 0.950 & 0.050 & 0.129 & 0.030 & 0.030 & 0.950 & 0.050 \\
LaOnline & 2.680 & 0.032 & 0.030 & 0.528 & 0.190 & 1.550 & 0.037 & 0.029 & 0.614 & 0.366 \\
DeCORER & 0.130 & 0.030 & 0.030 & 0.947 & 0.050 & 0.130 & 0.030 & 0.029 & 0.940 & 0.054 \\

\midrule
\multicolumn{11}{c}{$\epsilon \sim t(3)$} \\
\midrule
      & \multicolumn{5}{c}{$n_t = 100$} & \multicolumn{5}{c}{$n_t = 200$} \\
\cmidrule(lr){2-6}\cmidrule(lr){7-11}
Oracle & 0.501 & 0.060 & 0.059 & 0.945 & $\diagup$ & 0.501 & 0.060 & 0.059 & 0.945 & $\diagup$ \\
DeOffline & 0.510 & 0.060 & 0.059 & 0.944 & 0.051 & 0.510 & 0.060 & 0.059 & 0.944 & 0.051 \\
LaOnline & 15.296 & 0.114 & 0.059 & 0.470 & 0.179 & 11.790 & 0.117 & 0.058 & 0.532 & 0.280 \\
DeCORER & 0.538 & 0.062 & 0.059 & 0.937 & 0.051 & 0.531 & 0.061 & 0.058 & 0.934 & 0.054 \\
\cmidrule(lr){2-6}\cmidrule(lr){7-11}
      & \multicolumn{5}{c}{$n_t = 500$} & \multicolumn{5}{c}{$n_t = 1000$} \\
\cmidrule(lr){2-6}\cmidrule(lr){7-11}
Oracle & 0.501 & 0.060 & 0.059 & 0.945 & $\diagup$ & 0.501 & 0.060 & 0.059 & 0.945 & $\diagup$ \\
DeOffline & 0.510 & 0.060 & 0.059 & 0.944 & 0.051 & 0.510 & 0.060 & 0.059 & 0.944 & 0.051 \\
LaOnline & 11.436 & 0.105 & 0.059 & 0.520 & 0.260 & 8.478 & 0.085 & 0.058 & 0.549 & 0.297 \\
DeCORER & 0.528 & 0.061 & 0.058 & 0.938 & 0.053 & 0.525 & 0.061 & 0.058 & 0.934 & 0.054 \\
\bottomrule
\end{tabular}
  \begin{tablenotes}
	\footnotesize
	\item  MSE are reported times $10^{-2}$.
\end{tablenotes}
  \label{tab:DeCOER-heter-fix-75}
\vspace{-5cm}
\end{table}

\begin{table}[htbp]

\footnotesize
  \centering
   \renewcommand{\arraystretch}{0.6}
  \renewcommand\tabcolsep{6.0pt}
  \caption{Performance of Homogeneous model, with fixed the batch sample size $n_t=200 (\tau=0.75)$}
\begin{tabular}{ccccccccccc}
\toprule
Method & MSE & SD & SE & CP & FPR & MSE & SD & SE & CP & FPR \\
\midrule
\multicolumn{11}{c}{$\epsilon \sim \mathcal{N}(0,1)$} \\
\midrule
      & \multicolumn{5}{c}{$b = 10$} & \multicolumn{5}{c}{$b = 20$} \\
\cmidrule(lr){2-6}\cmidrule(lr){7-11}
Oracle & 0.058 & 0.024 & 0.024 & 0.955 & $\diagup$ & 0.028 & 0.017 & 0.017 & 0.944 & $\diagup$ \\
DeOffline & 0.059 & 0.024 & 0.024 & 0.954 & 0.051 & 0.028 & 0.017 & 0.017 & 0.943 & 0.050 \\
LaOnline & 0.190 & 0.024 & 0.023 & 0.640 & 0.141 & 0.094 & 0.017 & 0.016 & 0.655 & 0.135 \\
DeCORER & 0.061 & 0.024 & 0.023 & 0.935 & 0.057 & 0.030 & 0.017 & 0.016 & 0.928 & 0.053 \\
\cmidrule(lr){2-6}\cmidrule(lr){7-11}
      & \multicolumn{5}{c}{$b = 50$} & \multicolumn{5}{c}{$b = 100$} \\
\cmidrule(lr){2-6}\cmidrule(lr){7-11}
Oracle & 0.011 & 0.011 & 0.011 & 0.947 & $\diagup$ & 0.005 & 0.007 & 0.007 & 0.952 & $\diagup$ \\
DeOffline & 0.011 & 0.011 & 0.011 & 0.948 & 0.048 & 0.005 & 0.007 & 0.007 & 0.953 & 0.049 \\
LaOnline & 0.035 & 0.011 & 0.010 & 0.679 & 0.135 & 0.017 & 0.007 & 0.007 & 0.690 & 0.137 \\
DeCORER & 0.012 & 0.011 & 0.010 & 0.943 & 0.050 & 0.005 & 0.007 & 0.007 & 0.953 & 0.051 \\

\midrule
\multicolumn{11}{c}{$\epsilon \sim t(3)$} \\
\midrule
      & \multicolumn{5}{c}{$b = 10$} & \multicolumn{5}{c}{$b = 20$} \\
\cmidrule(lr){2-6}\cmidrule(lr){7-11}
Oracle & 0.219 & 0.047 & 0.046 & 0.947 & $\diagup$ & 0.108 & 0.033 & 0.033 & 0.958 & $\diagup$ \\
DeOffline & 0.217 & 0.047 & 0.045 & 0.940 & 0.051 & 0.109 & 0.033 & 0.032 & 0.953 & 0.049 \\
LaOnline & 0.692 & 0.052 & 0.042 & 0.648 & 0.213 & 0.316 & 0.036 & 0.031 & 0.677 & 0.235 \\
DeCORER & 0.228 & 0.047 & 0.042 & 0.915 & 0.063 & 0.112 & 0.033 & 0.031 & 0.936 & 0.058 \\
\cmidrule(lr){2-6}\cmidrule(lr){7-11}
      & \multicolumn{5}{c}{$b = 50$} & \multicolumn{5}{c}{$b = 100$} \\
\cmidrule(lr){2-6}\cmidrule(lr){7-11}
Oracle & 0.045 & 0.021 & 0.021 & 0.948 & $\diagup$ & 0.023 & 0.015 & 0.015 & 0.942 & $\diagup$ \\
DeOffline & 0.045 & 0.021 & 0.021 & 0.949 & 0.050 & 0.023 & 0.015 & 0.015 & 0.940 & 0.051 \\
LaOnline & 0.125 & 0.023 & 0.020 & 0.700 & 0.257 & 0.064 & 0.017 & 0.014 & 0.700 & 0.255 \\
DeCORER & 0.047 & 0.021 & 0.020 & 0.933 & 0.055 & 0.024 & 0.015 & 0.014 & 0.933 & 0.054 \\
\bottomrule
\end{tabular}
  \begin{tablenotes}
	\footnotesize
	\item  MSE are reported times $10^{-2}$.
\end{tablenotes}
  \label{tab:DeCOER-homo-stream-75}
\vspace{12pt}
  \centering
  \renewcommand{\arraystretch}{0.6}
  \caption{Performance of Heterogeneous model, with fixed the batch sample size $n_t=200 (\tau=0.75)$}
   \renewcommand{\arraystretch}{0.6}
  \renewcommand\tabcolsep{6.0pt}
\begin{tabular}{ccccccccccc}
\toprule
Method & MSE & SD & SE & CP & FPR & MSE & SD & SE & CP & FPR \\
\midrule
\multicolumn{11}{c}{$\epsilon \sim \mathcal{N}(0,1)$} \\
\midrule
      & \multicolumn{5}{c}{$b = 10$} & \multicolumn{5}{c}{$b = 20$} \\
\cmidrule(lr){2-6}\cmidrule(lr){7-11}
Oracle & 0.626 & 0.067 & 0.066 & 0.956 & $\diagup$ & 0.287 & 0.046 & 0.047 & 0.945 & $\diagup$ \\
DeOffline & 0.709 & 0.068 & 0.068 & 0.945 & 0.053 & 0.306 & 0.047 & 0.048 & 0.941 & 0.051 \\
LaOnline & 18.120 & 0.114 & 0.068 & 0.486 & 0.232 & 8.732 & 0.082 & 0.048 & 0.494 & 0.218 \\
DeCORER & 0.775 & 0.069 & 0.067 & 0.935 & 0.059 & 0.315 & 0.047 & 0.047 & 0.932 & 0.054 \\
\cmidrule(lr){2-6}\cmidrule(lr){7-11}
      & \multicolumn{5}{c}{$b = 50$} & \multicolumn{5}{c}{$b = 100$} \\
\cmidrule(lr){2-6}\cmidrule(lr){7-11}
Oracle & 0.126 & 0.030 & 0.030 & 0.949 & $\diagup$ & 0.054 & 0.020 & 0.021 & 0.955 & $\diagup$ \\
DeOffline & 0.129 & 0.030 & 0.030 & 0.950 & 0.050 & 0.055 & 0.020 & 0.021 & 0.953 & 0.049 \\
LaOnline & 3.229 & 0.050 & 0.030 & 0.502 & 0.218 & 1.727 & 0.035 & 0.021 & 0.495 & 0.204 \\
DeCORER & 0.130 & 0.030 & 0.030 & 0.947 & 0.050 & 0.056 & 0.020 & 0.021 & 0.954 & 0.049 \\

\midrule
\multicolumn{11}{c}{$\epsilon \sim t(3)$} \\
\midrule
      & \multicolumn{5}{c}{$b = 10$} & \multicolumn{5}{c}{$b = 20$} \\
\cmidrule(lr){2-6}\cmidrule(lr){7-11}
Oracle & 2.434 & 0.132 & 0.129 & 0.943 & $\diagup$ & 1.165 & 0.092 & 0.092 & 0.959 & $\diagup$ \\
DeOffline & 2.754 & 0.133 & 0.123 & 0.926 & 0.059 & 1.208 & 0.093 & 0.090 & 0.948 & 0.055 \\
LaOnline & 58.556 & 0.234 & 0.121 & 0.490 & 0.257 & 22.112 & 0.141 & 0.088 & 0.534 & 0.299 \\
DeCORER & 2.990 & 0.135 & 0.121 & 0.907 & 0.065 & 1.250 & 0.093 & 0.088 & 0.934 & 0.060 \\
\cmidrule(lr){2-6}\cmidrule(lr){7-11}
      & \multicolumn{5}{c}{$b = 50$} & \multicolumn{5}{c}{$b = 100$} \\
\cmidrule(lr){2-6}\cmidrule(lr){7-11}
Oracle & 0.501 & 0.060 & 0.059 & 0.945 & $\diagup$ & 0.225 & 0.041 & 0.041 & 0.947 & $\diagup$ \\
DeOffline & 0.510 & 0.060 & 0.059 & 0.944 & 0.051 & 0.230 & 0.041 & 0.041 & 0.943 & 0.052 \\
LaOnline & 11.790 & 0.117 & 0.058 & 0.532 & 0.280 & 5.192 & 0.070 & 0.041 & 0.533 & 0.279 \\
DeCORER & 0.531 & 0.061 & 0.058 & 0.934 & 0.054 & 0.236 & 0.042 & 0.041 & 0.937 & 0.054 \\
\bottomrule
\end{tabular}
  \begin{tablenotes}
	\footnotesize
	\item  MSE are reported times $10^{-2}$.
\end{tablenotes}
  \label{tab:DeCOER-heter-stream-75}
\end{table}
\end{document}